\documentclass[useAMS,usenatbib,babel]{mnras}

\usepackage[english]{babel}
\usepackage{amsmath}
\usepackage{balance}
\usepackage{amssymb,amsfonts,textcomp}
\usepackage{array}
\usepackage{supertabular}
\usepackage{slashbox}
\usepackage{hhline}
\usepackage{hyperref}  
\hypersetup{colorlinks=true, linkcolor=blue, citecolor=blue, filecolor=black, urlcolor=black}

\usepackage{flushend}
\usepackage{textcomp}
\usepackage{times}

\usepackage{graphicx}

\usepackage{dcolumn}
\usepackage{bm}
\usepackage{comment}
\usepackage{color}

\usepackage{slashbox}

\DeclareGraphicsExtensions{.jpg,.mps,.pdf,.png}

\newcommand{\mR}{{\cal{R}}}

\newcommand{\mP}{{\cal P}}
\newcommand{\mC}{{\cal C}}

\newcommand{\vx}{\textbf{x}}
\newcommand{\vr}{\textbf{r}}

\newcommand{\vk}{\textbf{k}}

\newcommand{\dd}{\hbox{d}}

\newcommand{\rmm}{{\rm m}}
\newcommand{\rmh}{{\rm h}}

\begin{document}

\author[C. Uhlemann,  C. Pichon, S. Codis, B.  L'Huillier, J. Kim et al. ]{
\parbox[t]{\textwidth}
{\hskip -0.0cm Cora Uhlemann$^{1,2,3}$, Christophe Pichon$^{4,5}$, Sandrine Codis$^{4,6}$, 
 Benjamin L'Huillier$^{7}$,\\
 Juhan Kim$^5$, Francis~Bernardeau$^{4,8}$, Changbom Park$^5$, Simon Prunet$^{4,9}$ }
\vspace*{6pt}\\
\noindent 
$^{1}$ Centre for Theoretical Cosmology, DAMTP, University of Cambridge, CB3 0WA, United Kingdom\\
$^{2}$ Fitzwilliam College, University of Cambridge, CB3 0DG, United Kingdom\\
$^{3}$ Institute for Theoretical Physics, Utrecht University, Princetonplein 5, 3584 CC Utrecht, The Netherlands\\
$^{4}$ CNRS \& UPMC, UMR 7095, Institut d'Astrophysique de Paris, F-75014, Paris, France\\
$^{5}$ Korea Institute for Advanced Study (KIAS), 85 Hoegiro, Dongdaemun-gu, Seoul, 02455, Republic of Korea\\
$^{6}$ Canadian Institute for Theoretical Astrophysics, University of Toronto, 60 St. George Street, Toronto, ON M5S 3H8, Canada\\
$^{7}$ Korea Astronomy and Space Science Institute (KASI), 776 Daedeokdae-ro, Yuseong-gu, Daejeon 34055, Republic of Korea\\
$^{8}$ CNRS \& CEA, UMR 3681, Institut de Physique Th\'eorique, F-91191 Gif-sur-Yvette, France\\
$^{9}$  Canada France Hawaii Telescope Corporation  65-1238 Mamalahoa Hwy  Kamuela, Hawaii 96743  USA
\\
}
\title[Non-Gaussian statistics of the projected cosmic density field]{
Cylinders out of a top hat: counts-in-cells for projected densities}

\maketitle
\begin{abstract}
{
Large deviation statistics  is implemented  to predict the statistics of cosmic densities in cylinders applicable to photometric surveys. It yields few percent accurate analytical predictions for the one-point probability distribution function (PDF) of densities in concentric  or  compensated cylinders; and also captures the density-dependence of their angular clustering (cylinder bias).  
All predictions are found to be in excellent agreement with the cosmological simulation Horizon Run 4  in the quasi-linear regime where standard perturbation theory normally breaks down. These results are combined with a simple local bias model that relates dark matter and tracer densities in cylinders and validated on simulated halo catalogues.  This formalism  can be used to probe cosmology with existing and upcoming photometric surveys like DES, Euclid or WFIRST containing billions of galaxies.
}
\end{abstract}

\section{Introduction}
Understanding the nature of dark energy is the main challenge cosmology is facing today, as it accounts for $\sim$70 per cent of the energy budget of our Universe. Multiplexed fiber  or slit-less instruments allow astronomers to 
collect thousands of spectroscopic redshifts (either ongoing GAMA \citep{Driver2016}, VIPERS \citep{VIPER}, or upcoming DESI \citep{DESI} PFS \citep{PFS}, MSE \citep{2016arXiv160600043M}). 
Such surveys provide a direct mean to probe accurately the cosmic evolution of the large-scale structure.
Yet  photometric redshift remains the most efficient method to map large volumes of the sky to position 
billions of galaxies.  Obtaining galaxy statistics such as the one-point density PDF and the angular correlation function in thick redshift slices does not require modelling the cosmology-dependent distance-redshift relation and allows us to accumulate enough statistics for populations of rare objects.
This has motivated international collaborations such as   DES \citep{DES2016}, Euclid \citep{Euclid}, WFIRST \citep{WFIRST}, LSST \citep{LSST}, KiDs \citep{KIDS},   
to carry out extensive large scale structure surveys
relying primarily on photometric redshifts to study the details of structure formation at different epochs and therefore offer insight into the engine of cosmic acceleration.

The prime estimators  which will be implemented on these surveys are angular clustering, weak lensing and supernovae. The former
will rely on photometric redshifts to estimate the distance of the lensed galaxies and improve the derived constraint on the equation of state of 
dark energy while  reconstructing the mass distribution of lenses on dozens  of  distinct planes simultaneously (implementing so called weak lensing tomography). The extracted catalogues  can at no extra cost be used to carry out count-in-cell statistics \citep{1993ApJ...417...36B,Efstathiou95,YangSaslaw11SDSS,2013MNRAS.435,BelMarinoni14,Bel16,FriedrichDES17,GruenDES17}.
Count-in-cell has been identified  as an ancillary probe  for these space missions, as it accesses   the non-linear regime of clustering analytically up to cosmic variance of the order of one. Indeed \cite{Uhlemann16log,Uhlemann17Kaiser}  derived the theory of one- and two-point counts-in-cells statistics for spherical symmetry in three dimensions using a fully analytical saddle-point approximation as a substitute for the numerical integration required  in \cite{Bernardeau14,Codis16correlations}. 
This analytic estimator can quantify  more modes of  the underlying clustering, hence has improved statistical power \citep{Codis2016DE}. It has also recently been extended to account for tracer bias so as to operate directly on galaxy catalogues \citep{Uhlemann17Bias}.

This paper theoretically models the one- and two-point statistics of densities-in-concentric-cylinders that can be obtained from measured 2D maps of the large scale structure in redshift bins. It focuses on extending and combining \cite{Uhlemann16log,Uhlemann17Bias} to  projected densities extracted from photometric -- quasi-cylindrical -- galaxy surveys with tracer bias. 
In a highly symmetric configuration such as cylindrical symmetry, one can take advantage of the fact that a non-linear solution to the gravitational dynamics (so-called cylindrical collapse) is known exactly and can be used to accurately probe the non-linear regime. In particular, this statistics allows us to study the density dependence of gravitational clustering in terms of the conditional one-point PDF of the density within underdense (resp. overdense) regions and the modulation of angular clustering therein. 

It also provides the basis for an application to cosmic shear experiments sensitive to the matter distribution itself, such as the weak lensing signal around shear peaks \citep{KacprzakDES16}, galaxy troughs \citep{Gruen16troughs} or more general density split statistics \citep{FriedrichDES17,GruenDES17}. {Indeed, density-split statistics from counts and lensing in cells can yield cosmological constraints competitive with the two-point function measurements if the stochasticity of the tracer-halo connection can be controlled \citep[see fig.~10 in][]{GruenDES17}}. In this respect, theoretical progress has been made in deriving the one-point PDF for weak lensing shear and local aperture mass in the mildly non-linear regime \citep{BV2000,Munshi00,Munshi01,Munshi04,Munshi14,ReimbergBernardeau17}. This is particularly interesting because, despite an approximate lognormality of 3D matter and tracers densities \citep{ColesJones91,Kayo01,Bel16,Hurtado-Gil17,Agrawal17} and weak lensing statistics \citep{Taruya02,Hilbert11,Clerkin17}, lognormal models have fundamental limitations regarding joint modelling of the two \citep{Xavier16}.

The outline of the paper is the following.  Section~\ref{sec:LDS}  presents how large deviation statistics can be used to obtain the PDF of dark matter densities in cylinders using cylindrical collapse dynamics and perturbative arguments.  Section~\ref{sec:validation}  compares these theoretical predictions for the one-point PDF and the two-point clustering of dark matter densities in cylinders to the Horizon Run 4 simulation.  
Such statistics allow us to probe differentially the slope of the density field in regions of low or high density.
Section~\ref{sec:haloes}  demonstrates how tracer densities can be related to dark matter densities using a simple local bias relation and 
 leads to accurate predictions for the statistics of tracers.  Section~\ref{sec:conclusion}  concludes and provides an outlook for future use of the results presented here.
 Appendix~\ref{sec:details} contains further details on the treatment in the main text, in particular regarding the cylinder depth and leading order cumulants from perturbation theory, while Appendix~\ref{app:lognormal} shortly addresses the accuracy of lognormal reconstructions.

\section{Large deviation  for projected densities}
\label{sec:LDS}

The general formalism of large deviation statistics for cosmic densities has been presented in \cite{Bernardeau2014,BernardeauReimberg15,Uhlemann16log} for 3D spherical cells. It is  presented here in the context of the 
density field smoothed in cylinders. In that case the general framework is left unchanged provided that 
\begin{enumerate}
\item The cylindrical instead of spherical filter is used to compute the initial correlation matrix of projected densities. 
\item The cylindrical instead of spherical collapse is used to map initial to final densities.
\item The dimension of the cylinders instead of spheres is properly rescaled from the initial to the final conditions according to 
mass conservation.
\item The finite character of the redshift slices is accounted for.
\end{enumerate}

After addressing those four points in the following subsection, Section \ref{sec:onepointPDF} describes how to compute the density PDF from large deviation statistics.

\subsection{Large deviation statistics with symmetric collapse models}
\label{sec:SC}
When considering a highly symmetric observable such as the density in spheres or infinitely long cylinders, one can argue that the most likely dynamics (amongst all possible mappings between the initial and final density fields) is the one respecting the symmetry \citep{2002A&A...382..412V}\footnote{This is a result of the so-called contraction principle in the context of large deviation theory as explained in \cite{BernardeauReimberg15},
which formalizes the idea that amongst all unlikely fates (in the tail of the PDF) the least unlikely one (symmetric collapse) dominates.}. For spherical and cylindrical symmetry, one can then take advantage of the fact that non-linear solutions to the gravitational dynamics are known explicitly in terms of the spherical and cylindrical collapse model, respectively. 

\subsubsection{Filters for cylinders and spheres}
To relate the average density in a region of a given shape to the underlying density field,  the appropriate top-hat filters are used.
For a finite cylinder with radius $R$ and depth $d$, a top-hat filter is used in cylindrical coordinates $W(r_{\perp},r_\|)\propto\Theta(R-r_\perp)\Theta(d/2-r_\|)$ where $(r_\perp,r_\|)$ is the distance perpendicular and parallel to the cylinder axis, while $\Theta$ is the Heaviside step function. The Fourier transform of the cylindrical filter is a product of a two- and one-dimensional top-hat filter
\begin{align}
\label{eq:filtercyl}
\tilde W_{\rm cyl}(Rk_\perp,dk_\|) &=\!\frac{2J_{1}(Rk_\perp)}{Rk_\perp} \text{sinc}\left(\frac{dk_\|}{2}\right)\,, \\
\notag &=\! \tilde W_{\rm 2D}(Rk_\perp)\tilde W_{\rm 1D}\left(\!\frac{dk_\|}{2}\!\right)\!,
\end{align}
where $J_1$ is the Bessel function of the first kind of order one. For an infinitely long cylinder $d\rightarrow \infty$, the cylindrical filter \eqref{eq:filtercyl} reduces to a disk top-hat filter $W_{\text{disk}}(r_{\rm 2D})\propto\Theta(R-r_{\rm 2D})$ in 2 dimensions. In comparison, for a spherical top-hat filter $W_{\rm{sph}}(r_{\rm 3D})\propto\Theta(R-r_{\rm 3D})$ in 3 dimensions one has
\begin{equation}
\tilde W_{\rm sph}(kR)=\frac{3(kR \cos(kR) - \sin(kR))}{(kR)^3}\,.
\label{eq:filter3D}
\end{equation}

For a galaxy redshift survey, one would eventually consider truncated conical filters $W_{\rm cone}(r_\|,\hat \gamma)$, that select objects depending on their radial distance $r_\|$ and solid angle $\hat \gamma$ on the sky. For conical shapes, one typically puts a selection function $F(r_\|)$ along the radial direction (pointing along the line of sight $\hat n_{\rm los}$) that encodes the mean density of observable objects at given distance (and hence redshift). Then one considers all solid angles $\hat \gamma$ around the line of sight $\hat n_{\rm los}$ with an angle $\theta=\angle(\hat n_{\rm los}, \hat \gamma)$ smaller than a given aperture $\theta_0$ such that $W_{\rm cone}(r_\|,\hat\gamma)\propto F(r_\|) \Theta(\theta_0- \theta)$. If the redshift slices are thin compared to the radial distance of the redshift bin, a truncated cone should be virtually identical to a cylinder centred at the corresponding mean redshift.

Using equations~(\ref{eq:filtercyl}) and (\ref{eq:filter3D}), the filtered linear covariance for concentric volumes is then obtained from the linear power spectrum according to 
\begin{align}
\label{eq:covmatrix}
\sigma^{2}_{{\rm s},L}(\mathcal R_{\rm s,1}\!,\mathcal R_{\rm s,2})&\!=\!\!\!\int\!\frac{\dd^{d_s}\vk}{(2\pi)^{d}}P_L(k)\tilde W_{\rm s}(\vk,\mR_{\rm s,1})\tilde W_{\rm s}(\vk,\mR_{\rm s,2})\,,
\end{align}
where $d_s$ encodes the dimension of integration and $\tilde W_{\rm s}$ is the Fourier transform of an appropriate filter enclosing a volume of a given shape $s$ (e.g cylinders, spheres) and characteristic dimensions $\mathcal R_{\rm s}$. For a cylinder,  $\mathcal R_{\rm cyl}=\{R,d\}$ and for a sphere $\mathcal R_{\rm sph}=R$. {Fig.~\ref{fig:scaledepvariance} shows a comparison of the scale-dependence of the variance for cylinders and spheres and Appendix~\ref{app:PSparam} describes an efficient power-law parametrisation for the covariance.}

\subsubsection{From spherical to cylindrical collapse}
Let us denote $\zeta_{\rm SC}(\tau)$ the non-linear transform of an initial fluctuation with linear density contrast $\tau$, in a $d_{s}$-dimensional sphere of radius  $R_{\rm ini}$, to the final density $\rho$ (in units of the average density) in a sphere of radius $R$ according to the spherical collapse model
\begin{equation}
\rho=\zeta_{\rm SC}(\tau)\,, 
\quad {\rm with}
\quad
\rho R^{d_{s}}= R_{\rm ini}^{d_{s}}\,, 
\label{eq:rho2tau}
\end{equation}
where the initial and final radii are connected through mass conservation. 
\cite{Bernardeau92,Bernardeau1995} showed that in 2D and 3D, the spherical collapse obeys the following differential equation
\begin{equation}
-\zeta_{\rm SC} \tau^{2} \zeta_{\rm SC}''+c (\tau \zeta_{\rm SC}')^{2}- \frac 3 2 \zeta_{\rm SC} \tau \zeta_{\rm SC}' +\frac 3 2 \zeta_{\rm SC}^{2} (\zeta_{\rm SC}-1)=0\,,
\label{eq:zeta-ED}
\end{equation}
with $c_{\rm 2D}=3/2$ and $c_{\rm 3D}=4/3$. While equation~\eqref{eq:zeta-ED} is strictly valid only for an Einstein-de Sitter background, it is also a good approximation for a general cosmological background. In both cases, one has $\zeta(\tau)\simeq 1+\tau+{\cal O}(\tau^{2})$ for small values of the linear density contrast $\tau$ as expected in the linear regime. The second and third orders then allows to predict the skewness from the spherical collapse dynamics as is described in Appendix~\ref{app:skewness}.
Equation~(\ref{eq:zeta-ED}) has a parametric solution for an Einstein-de Sitter background in three dimensions and can be solved numerically in other cases.
An explicit possible fit for the relation between the initial density contrast $\tau$ and the final density $\rho$ is given by
\begin{equation}
\rho_{\rm SC}(\tau)=\left (1-\tau/\nu\right )^{-\nu} \, \Leftrightarrow\,
\tau_{\rm SC}(\rho)=\nu(1-\rho^{-1/\nu})\,,
\label{eq:spherical-collapse}
\end{equation}
where $\nu$ depends on the dimension and the symmetry of collapse ($d_{s}=3$ for spherical and $d_{s}=2$ for cylindrical/disk-like) and can be adjusted to the actual values of the cosmological parameters. The Zeldovich approximation, which is exact in 1D, corresponds to setting $\nu_{\rm ZA}=1$\footnote{{For an analytical treatment of the PDF arising from 1D gravitational collapse and non-perturbative effects therein, we refer to \cite{vanderWoude017} which also addresses the relation of the PDF in Lagrangian and Eulerian space.}}.

In 3D, $\nu_{\rm 3D}=21/13\simeq 1.6$ provides a good description of the spherical dynamics for an Einstein-de Sitter background for the range of $\tau$ values of interest. In particular, this number allows us to reproduce the tree-order reduced skewness $S_3$ which captures the behaviour for small and intermediate values of $\tau$.
From equation~(\ref{eq:zeta-ED}), \cite{Bernardeau1995} found that for the 2D case asymptotically $\zeta(\tau)\propto \tau^{-(\sqrt{13}-1)/2}$ for large $\tau$ and therefore proposed to use the parametrisation~(\ref{eq:spherical-collapse}) with $\nu_{\rm 2D}=(\sqrt{13}-1)/2\sim 1.30$. If instead, in analogy to the 3D case, one wants to reproduce the value of the tree-order unsmoothed skewness for cylindrical symmetry $S_{3}^{\rm 2D}=36/7$ then one shall set $\nu_{\rm 2D}=1.4$. { Appendix~\ref{app:skewness}  recaps the expression for the skewness at tree-order in perturbation theory and its relation to the spherical collapse parameter $\nu$.}

\subsubsection{Relating initial to final volumes}
Because of  mass conservation, the dependence of the variance on the initial shape of the cells $\mR_{\rm ini}$ translates into a density dependence when the final dimensions of the cylinder $\mR=\{R,d\}$ are fixed. Hence, a prescription of the form $\mR_{\rm ini}=\mR_{\rm ini}(\mR,\rho)$ is needed to map the final to the initial dimensions of the cell given the enclosed final density. 
For a maximally symmetric case, like spherical symmetry in 3D, mass conservation completely fixes the relation between initial and final radii via $R_{\rm sph}^{\rm ini}=R_{\rm sph}\rho_{\rm sph}^{1/3}$. The same holds true for an infinitely long cylinder that corresponds to a 2D disk, for which mass conservation enforces to map the initial radius to the final one via 
\begin{align}
\label{eq:mapdisk}
\text{long cylinders: \quad} R_{\rm disk}^{\rm ini}=R_{\rm disk}\cdot\rho_{\rm disk}^{1/2}\,,
\end{align}
where $\rho_{\rm disk}$ is the surface density in the cylinder. 
In these maximally symmetric cases, the only unknown is the initial radius and the sole constraint of mass conservation is sufficient to fix it.

Conversely, for a finite cylinder, two unknowns have to be determined (radius and depth) from one single constraint that relates the initial and final volume according to mass conservation $V_{\rm cyl}^{\rm ini}= V_{\rm cyl} \rho_{\rm cyl}$. One can then distribute the density $\rho_{\rm cyl}$ between the two scales of the problem: the radius $R$ and depth $d$ of the cylinder. For a cylinder which has sphere-like dimensions $d\simeq 2R_{\rm 2D}$, one can expect that the ratio of cylinder radius to depth remains constant since the cylinder should collapse similarly to an equivalent sphere. Hence, for a sphere-like cylinder one should perform an equal splitting
\begin{align}
\label{eq:mapsphere}
\text{sphere-like cylinders: \quad} (R_{\rm cyl}^{\rm ini},d_{\rm cyl}^{\rm ini})=(R_{\rm cyl},d_{\rm cyl})\cdot \rho_{\rm cyl}^{1/3} \,,
\end{align}
as opposed to infinitely long cylinders for which all the density factor is attributed to the radius.

{In the remainder of the text, the focus will be on long cylinders with axis ratios $2R/d<1/10$ for which the approximation of infinitely long cylinder should hold.\footnote{It was checked that as far as the one-point PDF is concerned, sphere-like cylinders with $R=10$, $d=20$ Mpc$/h$ are well described by spherical collapse with $\nu=1.6$ and a sphere-like mapping of the axis ratio as in \eqref{eq:mapsphere}.} Indeed, for those axis ratios, previous studies for prolate spheroids (elongated symmetric ellipsoids that should closely resemble cylinders) in \cite{Yoshisato98,Yoshisato06} suggest that effective two-dimensional spherical collapse is accurate and the evolution of the axis corresponds to the one of infinitely long cylinders \eqref{eq:mapdisk}. {This  argument is seconded in Appendix~\ref{app:skewness} by comparing the skewness of the measured density field to tree-order perturbation theory.}

\subsection{The one-point density PDF in the large deviation regime}
\label{sec:onepointPDF}

Let us now compute the one-point PDF of the density field smoothed in a cylinder from large deviation statistics.

\subsubsection{PDF and decay-rate function for an initial Gaussian field}
The principles of large deviation statistics yield a formula for the PDF of finding a certain density in a symmetric volume given the initial conditions once it is assumed that the most likely dynamics that relates the initial to the final configurations is given by the appropriate collapse model. To achieve this goal, the main ingredient is the decay-rate function which encodes the exponential decay of the PDF. This formalism can be applied to any number of concentric symmetric volumes (e.g spheres, cylinders) as the symmetry is preserved in this configuration. For Gaussian initial conditions, which is assumed here\footnote{{See \cite{Uhlemann17nonG} for an extension of the formalism to include primordial non-Gaussianity.}}, the initial PDF of density contrasts $\tau_k$ in $N$ concentric cylinders with characteristic dimensions $\mR_k=\{R_k,d\}$ can be written as
\begin{equation}
 \label{eq:saddlePDFlogN-cell-ini}
\mP_{\{\mathcal R_k\}}^{\rm{ini}}(\{\tau_k\})\!=\! \sqrt{\det\!\!\left[\frac{\partial^{2}\Psi^{\rm ini}_{\{\mathcal R_k\}}}{\partial \tau_{i}\partial \tau_{j}}\!\right]} \frac{\! \exp\left[-\Psi^{\rm ini}_{\{\mathcal R_k\}}(\{\tau_k\})\right]}{ (2 \pi)^{N/2}} 
\,,
 \end{equation}
where $\mathcal R^{\rm ini}_k=(R^{\rm ini}_k,d^{\rm ini}_k)$ encodes the characteristic dimensions of the initial cylinders. 
The  initial decay-rate function is  given by the usual quadratic form in the initial density contrasts $\tau_k$ in cylinders with characteristic dimensions $\mathcal R_k$ 
\begin{align}
\Psi^{\rm ini}_{\{\mathcal R_k\}}(\{ \tau_{k}\})=\frac{1}{2}\sum_{i,j}\Xi_{ij}(\{ \mathcal R^{\rm ini}_k \})\,\tau_{i}\tau_{j}\,,
\label{PsiDefIni}
\end{align}
where $\Xi_{ij}$ is the inverse of the linear covariance matrix, $\sigma_L^{2}(\mathcal R^{\rm ini}_{i},\mathcal R^{\rm ini}_{j})$ obtained from equation~\eqref{eq:covmatrix} and encoding all dependency with respect to the initial power spectrum and the volume which is averaged over. Note that equation~\eqref{eq:saddlePDFlogN-cell-ini} is merely a rewriting of a Gaussian distribution, emphasizing the central role of the decay-rate function~\eqref{PsiDefIni}.

\subsubsection{Large deviation statistics PDF for an evolved field}  

Using the so-called contraction principle in the zero variance limit \citep{BernardeauReimberg15}
\begin{equation}
\Psi(Y)=\inf_{X\rightarrow Y} \Psi(X)
\end{equation}
the final decay-rate function is obtained from re-expressing the initial decay-rate function in terms of the final densities $\rho$ and characteristic dimensions of the cylinder $\mR=\{R,d\}$ 
\begin{align}
\Psi_{\{\mR_k\}}(\!\{ \rho_{k}\!\})\!=\!\frac{1}{2}\!\sum_{i,j}\!\Xi_{ij}(\{\mR_k^{\rm ini}\!(\mR_k,\rho_k)\! \})\tau_{\rm SC}(\rho_{i})\tau_{\rm SC}(\rho_{j})\,,
\label{PsiDef}
\end{align}
where the mapping between initial and final densities $\tau_{\rm SC}(\rho)$ is given by the cylindrical collapse from equation~\eqref{eq:spherical-collapse}. The non-linear covariance is approximated by rescaling the linear covariance from equation~\eqref{eq:covmatrix} using a reference scale 
\begin{align}
\label{eq:variancerescaling}
\sigma(\mR_{\rm ini})\equiv \sigma(\mR_{\rm ini},\mR_{\rm ini}) \simeq \frac{\sigma(\mR)}{\sigma_{L}(\mR)} \sigma_{L}(\mR_{\rm ini}) \,,
\end{align}
while the relation between the initial and final characteristic dimensions of the cylinders $\mR^{\rm ini}(\mR,\rho)$ is given by equations~\eqref{eq:mapdisk} for long cylinders. {The non-linear variance at reference scale -- also known as driving parameter in the context of large deviations theory --  can be either measured directly from the densities in spheres extracted from the simulation, derived following equation~\eqref{eq:covmatrix} using the non-linear power spectrum or found by a best fit to the measured one-point PDF \citep{Codis2016DE}.} 

The PDF of evolved densities in concentric cylinders, $\mP_{\{\mR_k\}}(\{\rho_k\})$,  at one point can then be obtained in full analogy to the case for densities in concentric spheres as discussed in \cite{Uhlemann16log}. Therein, it has been shown that the saddle-point technique provides an excellent approximation to the exact result from large-deviations statistics if a suitable variable is chosen. {For a single cylinder the suitable variable is the log-density $\mu=\log\rho$ while for concentric cylinders one should use logarithmically mapped combinations of the densities, see Section~\ref{subsec:concentric} for the case of two cylinders}. The PDF of those mapped densities is then given by
\begin{subequations}
 \label{eq:saddlePDFN-cell}
 \begin{align}
 \label{eq:saddlePDFlogN-cell}
\mP_{\mu,\{\mR_k\}}^{\text{cyl}}(\{\mu_k\})&= \sqrt{\det\left[\frac{\partial^{2}\Psi_{\{\mR_k\}}}{\partial \mu_{i}\partial \mu_{j}}\right]} \frac{ \exp\left[-\Psi_{\{\mR_k\}}\right]}{ (2 \pi)^{N/2}} 
\,,
 \end{align}
where $\Psi$ is now given by equation~\eqref{PsiDef}. The PDF of the log-density can be related to the PDF of the density via a simple change of variables
\begin{align}
\mP_{\{\mR_k\}}^{\text{cyl}}(\{\rho_k\}) &=\mP_{\mu,\{\mR_k\}}^{\text{cyl}}[\{\mu_k(\{\rho_i\})\}] \left|\det\left[\frac{\partial\mu_{i}}{\partial \rho_{j}}\right]\right|  \,.
\label{eq:Prhofrommu}
\end{align}
Equation~(\ref{eq:saddlePDFlogN-cell}) assumes that the mean of $\mu_{j}$ vanishes independently of the variance. For a generic non-linear mapping, it will translate into a mean density which can deviate from one as $\sigma$ grows. In order to avoid this effect, one has to consider the shifted PDF 
\begin{equation}
\label{eq:saddlePDFN-cellnorm}
\hat\mP_{\mu,\{\mR_k\}}^{\text{cyl}}(\{\mu_k\})=\mP_{\mu,\{\mR_k\}}^{\text{cyl}}(\{\tilde\mu_k=\mu_k-\left\langle\mu_{k}\right\rangle\})\,,
\end{equation}
with the shifts $\left\langle\mu_{k}\right\rangle$ chosen such that the resulting mean densities are one $\langle\rho_i\rangle=1\,\forall i=1,\cdots,n$. Furthermore, since the saddle-point method yields only an approximation to the exact PDF, the PDF obtained from equation~\eqref{eq:saddlePDFN-cell} is not necessarily perfectly normalized \citep{Uhlemann17Kaiser}. In practice, this can be corrected for by considering
\begin{equation}
\label{eq:saddlePDFN-cellnorm2}
\hat\mP_{\mR}^{\text{cyl}}(\{\rho_k\})=\mP_{\mR}^{\text{cyl}}(\{\rho_k\})/\langle1\rangle\,,
\end{equation}
with the shorthand notation $\langle 1\rangle= \prod_k \int_0^\infty \dd\rho_k\, \mP_{\mR}^{\text{cyl}}(\{\rho_k\})$.
\end{subequations}
{Note that this effect is however very minor as the deviation of $\langle 1\rangle$ from unity is typically at the sub-percent level.}

\subsection{The two-point cylinder bias in the large deviation regime}
Let us finally close this section by turning to the two-point statistics of densities in cylinders and in particular to the two-point cylinder bias from large deviation statistics.
\subsubsection{Kaiser bias for an initial Gaussian field}
The initial two-point bias used in previous works on densities-in-spheres \citep[see e.g.][]{Bernardeau96bias,Codis16correlations,Uhlemann17Kaiser} can also be straightforwardly generalised to cylinders. Cylinder bias is defined as the ratio between the conditional mean density contrast induced by a given initial density contrast $\tau_0$ in a cylinder of characteristic dimensions $\mR_{\rm ini}$ at separation $r_\perp$ (perpendicular to the cylinder axis) and the average correlation
\begin{equation}
\label{eq:Kaiserbiasini}
b_{\rm ini}(\tau_{0},r_\perp)= \frac{\langle\tau(\vx+\vr_\perp)|\tau(\vx)\equiv \tau_{0}\rangle}{\langle\tau(\vx+\vr_\perp)\tau(\vx)\rangle} = \frac{\displaystyle\! \int \! \dd\tau \,\tau\,\mP^{\rm 2pt}_{\mR_{\rm ini}}(\tau_{0},\tau;r_\perp)}{\mP_{\mR_{\rm ini}}(\tau_{0}) \xi_{L}({\mR_{\rm ini}},r_\perp)} \,, \notag
\end{equation}
and can be expressed in terms of the joint PDF $\mP^{\rm 2pt}_{\mR_{\rm ini}}$ of densities in cylinders of equal characteristic dimensions  $\mR_{\rm ini}$ at separation $r_\perp$ and their correlation function $\xi(\mR_{\rm ini},r_\perp)=\langle\tau(\vx+\vr_\perp)\tau(\vx)\rangle$. The initial correlation function is obtained from the linear power spectrum as 
\begin{align}
\label{eq:corrlin}
\xi_L(\mR_{\rm ini},r_\perp) &\!=\! \int \! \! \frac{\dd^3 \vk}{(2\pi)^3} P_L(k)\tilde W^2(\vk,\mR_{\rm ini}) \exp(i \vk_\perp \cdot \vr_\perp\!).
\end{align} 
For Gaussian initial conditions, it is straightforward to determine and diagonalise the linear covariance matrix to find that the cylinder bias predicted by large deviation theory is independent of separation $r_\perp$ and reads
\begin{align}
b_{\rm ini}(\tau,r_\perp)\equiv b_{\rm ini}(\tau)
=\frac{\tau}{\sigma_L^{2}(\mR_{\rm ini})}\,,
\label{eq:biasG}
\end{align}
which is proportional to the initial overdensity $\tau$ as expected from \cite{Kaiser84}. Since the cylinder bias is independent of the separation $r_\perp$, the density dependence of the two-point statistics of densities in cylinders can be factorised at large separations $r_\perp\gg 2R$ according to
\begin{align}
\label{eq:jointPDFlargesep}
\frac{\mP_{\mR_{\rm ini}}^{\rm 2pt}(\tau,\tau';r_\perp)}{\mP_{\mR_{\rm ini}}(\tau)\mP_{\mR_{\rm ini}}(\tau')}\!\simeq\! 1\!+\!\xi_L (\mR_{\rm ini},r_\perp) b_{\rm ini}(\tau)b_{\rm ini}(\tau')\,.
\end{align}
\subsubsection{Large deviation statistics bias for an evolved field}
The cylinder bias can be defined in the evolved density field in complete analogy to the initial one from equation~\eqref{eq:Kaiserbiasini} and reads
\begin{equation}
\label{eq:defdensitybias}
b^{\text{cyl}}_{\mR}(\rho_{0},r_\perp)= \frac{\langle\rho(\vx+\vr_\perp)|\rho(\vx)\equiv \rho_{0}\rangle-1}{\langle\rho(\vx+\vr_\perp)\rho(\vx)\rangle-1} \,.
\end{equation}
Following the same prescription as for the PDF for replacing initial and final densities and characteristic dimensions of the cylinder,  large deviation theory predicts that the final cylinder bias is independent of the transverse separation and given by
\begin{subequations}
\label{eq:densitybias}
\begin{align}
\label{eq:densitybias2}
b_{\mR}^{\text{cyl}}(\rho) &= \frac{\sigma_{L}^2(\mR)}{\sigma_\mu^2(\mR)}\frac{\tau_{\rm SC}(\rho) }{\sigma_{L}^2(\mR_{\rm ini}(\mR,\rho))} \,.
\end{align}
Equation~(\ref{eq:densitybias2}) must also  be shifted and normalised \citep{Uhlemann17Kaiser} according to
\begin{align}
\hat b_{\mR}^{\text{cyl}}(\rho) &= \frac{b_{\mR}^{\text{cyl}}(\rho)- \langle b_{\mR}^{\text{cyl}}(\rho)\rangle}{\langle\rho  b_{\mR}^{\text{cyl}}(\rho)\rangle- \langle b_{\mR}^{\text{cyl}}(\rho)\rangle}\,. \label{eq:biasnorm1cell}
\end{align}
\end{subequations}
where the averages denoted by $\langle\cdot\rangle$ are computed as integrals with the one-point PDF from equation~\eqref{eq:saddlePDFN-cell}. As before, the prediction depends on the assumed mapping between initial and final radii and depths and the choice of the filter function for obtaining the linear variance $\sigma_{L}$.

\section{Validation for dark matter densities}
\label{sec:validation}
Let us now  compare the theoretical predictions for the one-point PDF and the two-point clustering of dark matter densities in cylinders to the cosmological N-body simulation Horizon Run 4.{To avoid confusion between dark matter densities considered here and tracer densities that will be used in Section~\ref{sec:haloes}, from now on  $\rho_\rmm$ stands for matter densities and $\rho_\rmh$ for halo densities.}

\subsection{Horizon Run 4 simulation}
The Horizon Run 4 simulation \citep[HR4,][]{HR4} is a massive $N$-body simulation, 
evolving $6300^3$ particles in a 3.15 Gpc$/h$ box  using the GOTPM TreePM code \citep{GOTPM}. 
It assumes a WMAP-5 cosmology, with
$(\Omega_\text{m},\Omega_\Lambda,\Omega_\text{b},h,\sigma_8,n_\text{s}) = (0.26,0.74,0.044,0.72,0.79, 0.96
)$, yielding a particle mass of $9\times 10^{9} h^{-1}\,M_\odot$.
The initial conditions were generated at $z=100$ using the second order
Lagrangian perturbation theory, which ensures accurate power spectrum
and halo mass function at redshift 0 \citep[see][ for details]{LHuillier14}.
For the purpose of this paper, cylinders of various radii and lengths were extracted from 
the dark matter and halo catalogues at fixed redshift and used to measure the statistical properties of their encompassed mean densities. {Error bars for the PDFs are determined from the standard error on the mean that is obtained by dividing the full simulation cube into 8 subsamples and computing the corresponding histograms.} A spherical slice of width 60 Mpc$/h$ was also extracted 
at redshift $0.36$ and projected onto the sky using the HealPix\protect\footnote{http://healpix.jpl.nasa.gov/} equal area scheme to produce
Fig.~\ref{fig:HR4lightconesphere}.
\begin{table}
\centering
  \renewcommand{\arraystretch}{1.2} 
\begin{tabular}{|c||c|c|c|c|c|c|c|}
\hline
$z=0.7$& \backslashbox{d}{R} & 3 & 5 & 7 &10  \\\hline\hline
$\hat \sigma_\rho$ & 150 & 0.330 & 0.274 & 0.242 & 0.209\\
$\hat \sigma_\mu$ & 150 & 0.300 & 0.259 & 0.232 & 0.203\\\hline
$\hat \sigma_\rho$ & 300 & 0.235 & 0.196 & 0.173 & 0.150 \\
$\hat \sigma_\mu$ & 300 & 0.22 & 0.190 & 0.169 & 0.148  \\\hline
$\hat \sigma_\mu$ & 450 & 0.193 & 0.161 & 0.142 & 0.123 \\
$\hat \sigma_\mu$ & 450 & 0.185 & 0.157 & 0.140 & 0.122\\\hline
$\hat \sigma_\rho$ & 700 & 0.154 & 0.129 & 0.114 & 0.099 \\
$\hat \sigma_\mu$ & 700 & 0.150 & 0.127 & 0.113 & 0.099\\\hline
\end{tabular}
\caption{Variances of the cylindrical density $\rho$ and log-density $\mu=\log\rho$ for different lenghts $d$ [Mpc/$h$] and radii $R$ [Mpc/$h$] at redshift $z=0.7$ as measured from the HR4 simulation.}
\label{tab:variance}
\end{table} 
\subsection{One-point PDF of density in single cylinders}
The explicit formula for the one-point PDF for dark matter densities $\rho_\rmm$ within a cylinder of radius $R$ and depth $d$ at redshift $z$ is expressed as
\begin{equation}
\hskip -0.1cm \mP_\mR^{\text{cyl}}(\rho_\rmm) \!=\! \sqrt{\frac{\Psi''_{R}(\rho_\rmm)+\Psi'_{R}(\rho_\rmm)/\rho_\rmm}{2\pi \sigma^{2}_\mu}} \exp\left(\!-\frac{\Psi_R(\rho_\rmm)}{\sigma^{2}_\mu}\!\right),
\label{eq:PDFfromPsi2}
\end{equation}
where the prime denotes a derivative with respect to $\rho_\rmm$ and 
\begin{equation}
\Psi_\mR(\rho_\rmm)= \frac{\tau^{2}_{\rm SC}(\rho_\rmm) \sigma_L^{2}(\mR)}{2\sigma_L^2(\mR_{\rm ini}(\mR,\rho_\rmm))}\,.
\label{eq:Psiquad}
\end{equation}
Note that $\sigma_\mu\equiv\sigma_\mu(\mR,z)$ is the non-linear variance of the log-density in a cylinder with characteristic dimensions $\mR=(R,d)$, which enters because the formula has been derived from an analytic approximation based on the log-density $\mu_\rmm=\log\rho_\rmm$, while $\sigma_L$ is the linear variance. Table~\ref{tab:variance} reports the values for the variance of the density and log-density for different radii and depths of the cylinders.
In Equation~(\ref{eq:Psiquad}), $\tau_{\rm SC}(\rho_\rmm)$ is the linear density contrast which can be mapped to the non-linearly evolved density $\rho_\rmm$ using the collapse model \eqref{eq:spherical-collapse} where the parameter $\nu=1.4$ characterises the dynamics of cylindrical collapse and matches the high-redshift skewness obtained from perturbation theory. The initial (Lagrangian) characteristic dimensions of the cylinder $\mR_{\rm ini}(\mR,\rho_\rmm)$ are mapped using equation~\eqref{eq:mapdisk} for long cylinders. To ensure a unit mean density and the correct normalization of the PDF, as described in Section~\ref{sec:onepointPDF} one can simply evaluate the PDF obtained from equation~\eqref{eq:PDFfromPsi2} according to
\begin{align}
\label{eq:PDFfromPsi2norm}
\hat\mP_\mR^{\text{cyl}}(\rho_\rmm)= \mP_\mR^{\text{cyl}}\left(\rho_\rmm \, \frac{\langle\rho_\rmm\rangle}{\langle 1\rangle}\right) \cdot \frac{\langle\rho_\rmm\rangle}{\langle 1\rangle^2} \,,
\end{align}
with the shorthand notation $\langle f(\rho_\rmm)\rangle=\int_0^\infty \dd\rho_\rmm\, f(\rho_\rmm)\mP_{\mR}^{\text{cyl}}(\rho_\rmm) $. 

Fig.~\ref{fig:saddlePDFvsHorizoncyl} compares the theoretical prediction for long cylinders with fixed depth $d=150$~Mpc$/h$ and various radii $R$ to the simulation results. The upper panel gives an overall view on the exponential decay of the one-point PDF for large deviations from the mean density. {It is complemented by the lower panel which shows the residuals around the maximum of the PDF, and displays the 2-$\sigma_{\mu}$ region around the mean for the log-density since the PDF of the log-density is close to Gaussian and hence allows for a fair comparison of under- and over-densities.} {An equivalent plot with fixed radius $R=5$~Mpc$/h$ and varying depth $d$ is shown in Fig.~\ref{fig:saddlePDFvsHorizonvaryD}.}
{Given the few percent-level agreement between the theoretical prediction and the simulation\footnote{Note that in contrast, the lognormal model shows residuals of about 5-20\%  in the central region and larger deviations in the tails, see Fig.~\ref{fig:lognormalPDFvsHorizonfixd} in Appendix~\ref{app:lognormal}.}, 
a future application of counts-in-cells statistics to photometric surveys is in order. The description is next generalised to concentric cylinders and two-point clustering that  allow to consider density-split statistics. To connect the results for dark matter to tracer densities, tracer bias is also discussed in Section~\ref{sec:haloes}  following closely \cite{Uhlemann17Bias} .}

\begin{figure}
\includegraphics[width=\columnwidth]{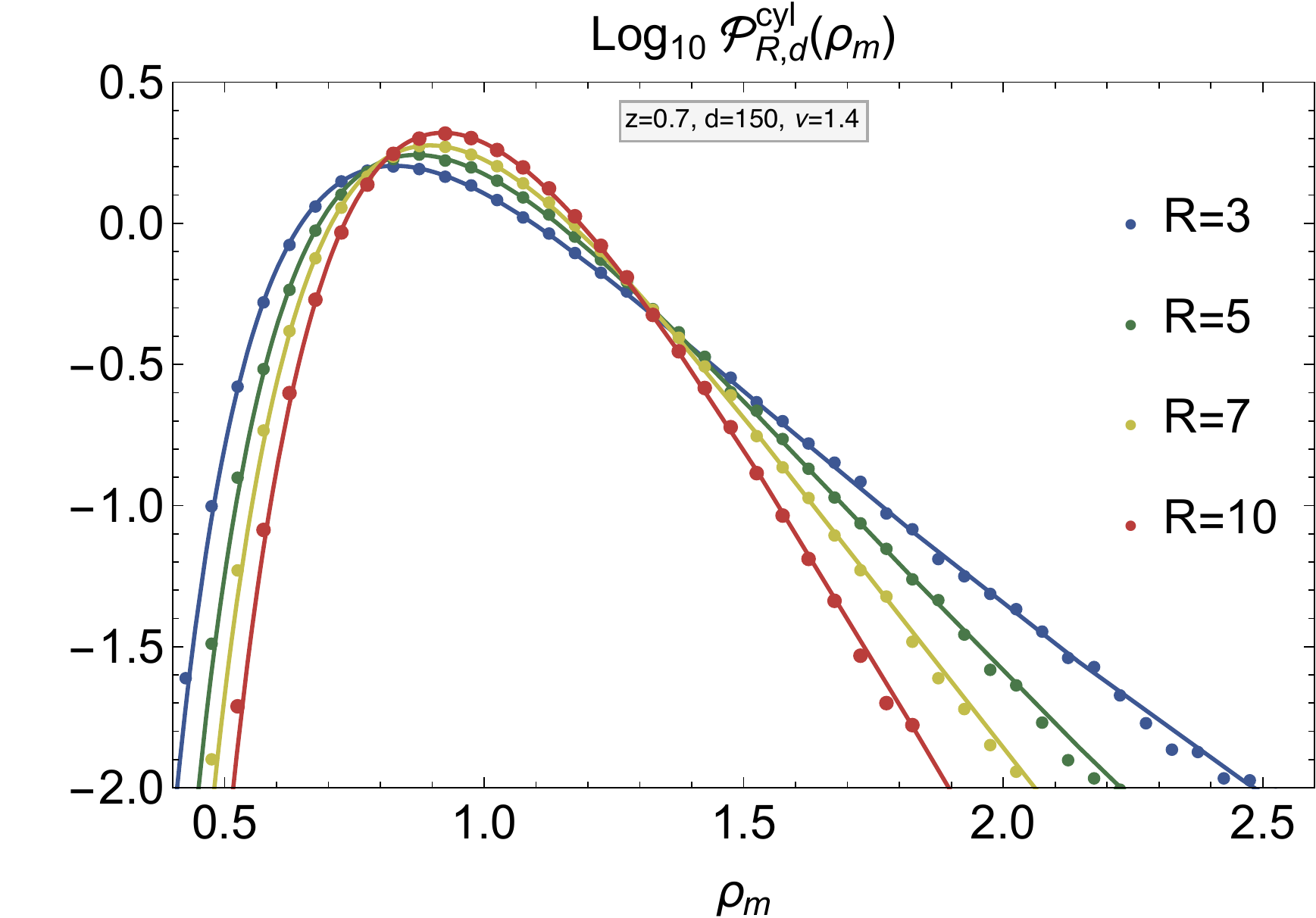}\\
\includegraphics[width=\columnwidth]{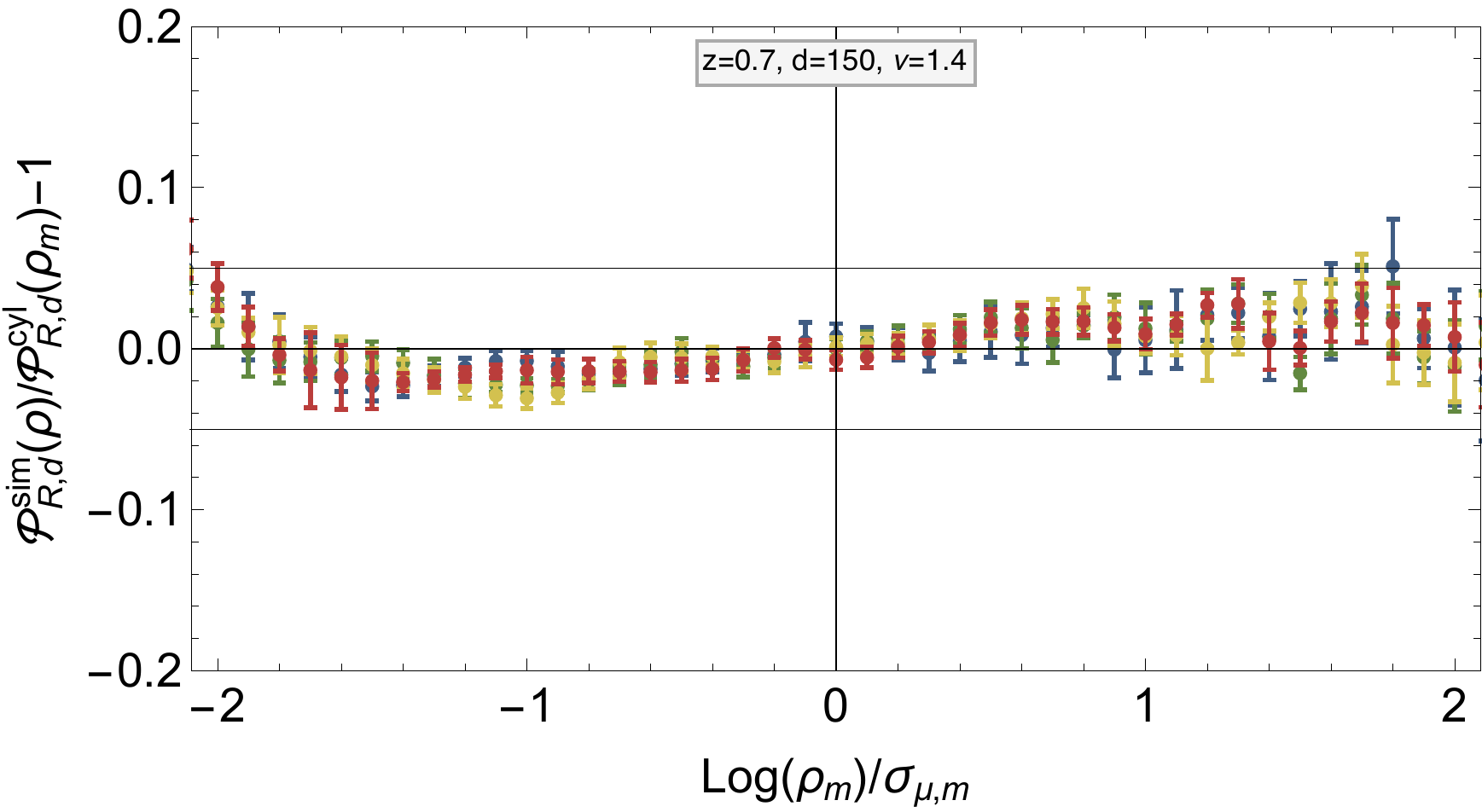}
\caption{{\it (Upper panel)} PDF of the matter density field smoothed in cylindrical cells of length $d=150$ Mpc/$h$ and radii $R=3,5,7,10$ Mpc/$h$ at redshift $z=0.7$. Shown are the HR4 measurements (data points) against the theoretical predictions computed with the finite cylinder filter given by equations~\eqref{eq:filtercyl}~and~\eqref{eq:mapdisk} together with the spherical collapse parametrized by $\nu=1.4$ (solid lines). {\it (Lower panel)} Residuals of the theoretical prediction using finite cylinder filter against the measurements around the maximum of the PDF for log density within two-sigma from the mean.}
   \label{fig:saddlePDFvsHorizoncyl}
\end{figure}

\subsection{One-point PDF of densities in concentric cylinders}
\label{subsec:concentric}

\subsubsection{Two concentric cylinders}
It is of interest to consider the joint PDF of concentric cylinders as it allows us to deduce conditionals which may e.g. act as proxies for peak or voids.
To obtain such PDF for the densities $\rho_{1},\rho_{2}$ in the cylinders of equal depth $d$ and radii $R_{1}<R_{2}$, one can also apply a logarithmic density mapping that relies on the sum and difference of mass in the two cylinders, \citep[see for the analogous sphere case][]{Uhlemann16log}
 \label{eq:eq:log-mass}
 \begin{align}
 \mu_{1}= \log  \left(r_{12}^{2}\rho_{2,\rmm}+\rho_{1,\rmm}\right)  \,,\,
   \mu_{2}= \log \left(r_{12}^{2}\rho_{2,\rmm}-\rho_{1,\rmm}\right)\,,
 \end{align}
where the relative shell thickness is $r_{12}=R_{2}/R_{1}>1$. The PDF $\mP_{R_1,R_2}^{\text{cyl}}(\rho_{1,\rmm},\rho_{2,\rmm})$ can then be approximated via equation~\eqref{eq:saddlePDFN-cell}. Analogously to the one cell case, one still has to enforce the mean and normalization for the saddle point PDF following the procedure described in equations~(\ref{eq:saddlePDFN-cellnorm})-(\ref{eq:saddlePDFN-cellnorm2}). 
It is convenient to express the joint density PDF $\mP_{R_1,R_2}^{\text{cyl}}(\rho_{1,\rmm},\rho_{2,\rmm})$ in terms of the inner density $\rho_{1,\rmm}$ and the density in the outer shell $\rho_{12,\rmm}= (R_2^2\rho_{2,\rmm}-R_1^2\rho_{1,\rmm})/(R_2^2-R_1^2)$.
Note that for the sake of simplicity,  a simpler scheme is used to compute the linear covariance matrix here, namely a power-law parametrisation of the power spectrum \eqref{eq:sigijparam} with index $n=n_{\text{cyl}}(R,d)$ according to equation~\eqref{eq:ncyl} instead of the exact linear power spectrum, see Appendix~\ref{app:PSparam}.

\begin{figure}
\includegraphics[width=\columnwidth]{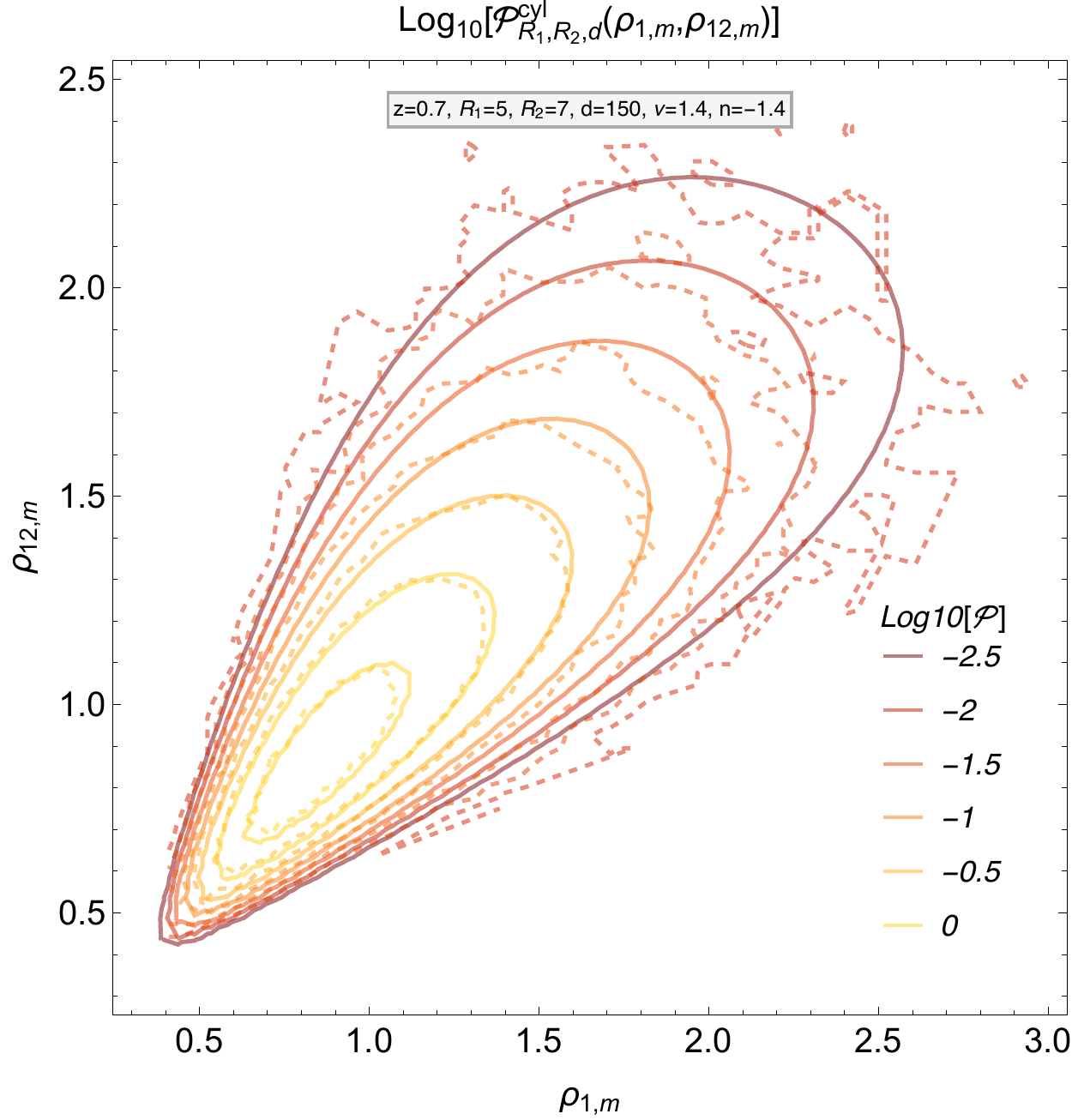}
\caption{Joint PDF of the projected matter densities in concentric cylinders of depth $d=150$ Mpc/$h$ and radii $R_1=5$ Mpc/$h$ and $R_2=7$ Mpc/$h$ at redshift $z=0.7$ from the saddle point approximation with $\nu=1.4$ and the disk filter for a power-law spectrum with $n=-1.4$ (solid lines) compared to the HR4 measurements (dashed lines).
}
\label{fig:saddlePDF2cellvsmeasurements}
\end{figure}
The result is displayed in Fig.~\ref{fig:saddlePDF2cellvsmeasurements} and compared to  measurements. Overall   a very good agreement is found.

\subsubsection{One-point PDF of densities in cylindrical rings}
From the joint PDF of densities in concentric cylinders, one can obtain the PDF  of a cylindrical shell and conditionals given inner over- or under-density via a simple marginalisation over the inner density
\begin{align}
\label{eq:PDFshell}
&\mP_{R_1\!,R_2}^{\text{shell}}(\rho_{12,\rmm})\!=\!\int\!\! \dd\rho_{1,\rmm} \mP_{R_1\!,R_2}^{\text{cyl}}(\rho_{1,\rmm},\rho_{12,\rmm})\,,\\
\label{eq:PDFshellcond}
&\mP_{R_1\!,R_2}^{\text{shell}}(\rho_{12,\rmm}|\rho_{1,\rmm}\!\gtrless\! 1)\\
&\quad = \frac{\displaystyle\int \dd\rho_{1,\rmm} \mP_{R_1\!,R_2}^{\text{cyl}}(\rho_{1,\rmm},\rho_{12,\rmm}) \Theta(\pm(\rho_{1,\rmm}\!-\!1))}{\displaystyle\!\! \int\! \dd\rho_{12,\rmm}\!\int\!\! \dd\rho_{1,\rmm} \mP_{R_1\!,R_2}^{\text{cyl}}(\rho_{1,\rmm},\rho_{12,\rmm}) \Theta(\pm(\rho_{1,\rmm}\!-\!1))}\,. \nonumber
\end{align}
A comparison between the marginal PDF and conditionals for over- and under-dense environments is shown in Fig.~\ref{fig:saddlePDFrho12condvsmeasurements}, while Fig.~\ref{fig:saddlePDFrho12vsmeasurements} focuses on the marginal PDF in  cylindrical rings. {For both statistics, very good agreement is found between the simulation measurements and the theoretical prediction, despite the simplifying assumption for the power spectrum. This is encouraging because} statistics conditional on over- or underdensities have recently gained renewed interest in the context of weak lensing around galaxy troughs \citep{Gruen16troughs} or more general density quantiles \citep{FriedrichDES17}. While  the statistics was here conditioned  on the underlying dark matter density, in practice one has to rely on tracer densities. To this end, the inclusion of tracer bias is discussed in Section~\ref{sec:haloes}. The marginal PDF of the densities in cylindrical rings gives a qualitative impression of the expected weak lensing signal that can probe the dark matter density up to a mass-sheet degeneracy, hence requires the use of compensated filters such as the difference of two cylinders. {Note that the relation between conical cells and long cylinders can change the underlying analytical properties of the cumulant generating functions \citep[see equation (34) in][]{BV2000}. However, large deviation statistics can be applied to many continuous nonlinear functionals of density profiles \citep{ReimbergBernardeau17}. Investigating approximations to compute PDFs for weak lensing convergence and cosmic shear fully analytically in the spirit of the log-density will be the subject of future work.}

\begin{figure}
\includegraphics[width=\columnwidth]{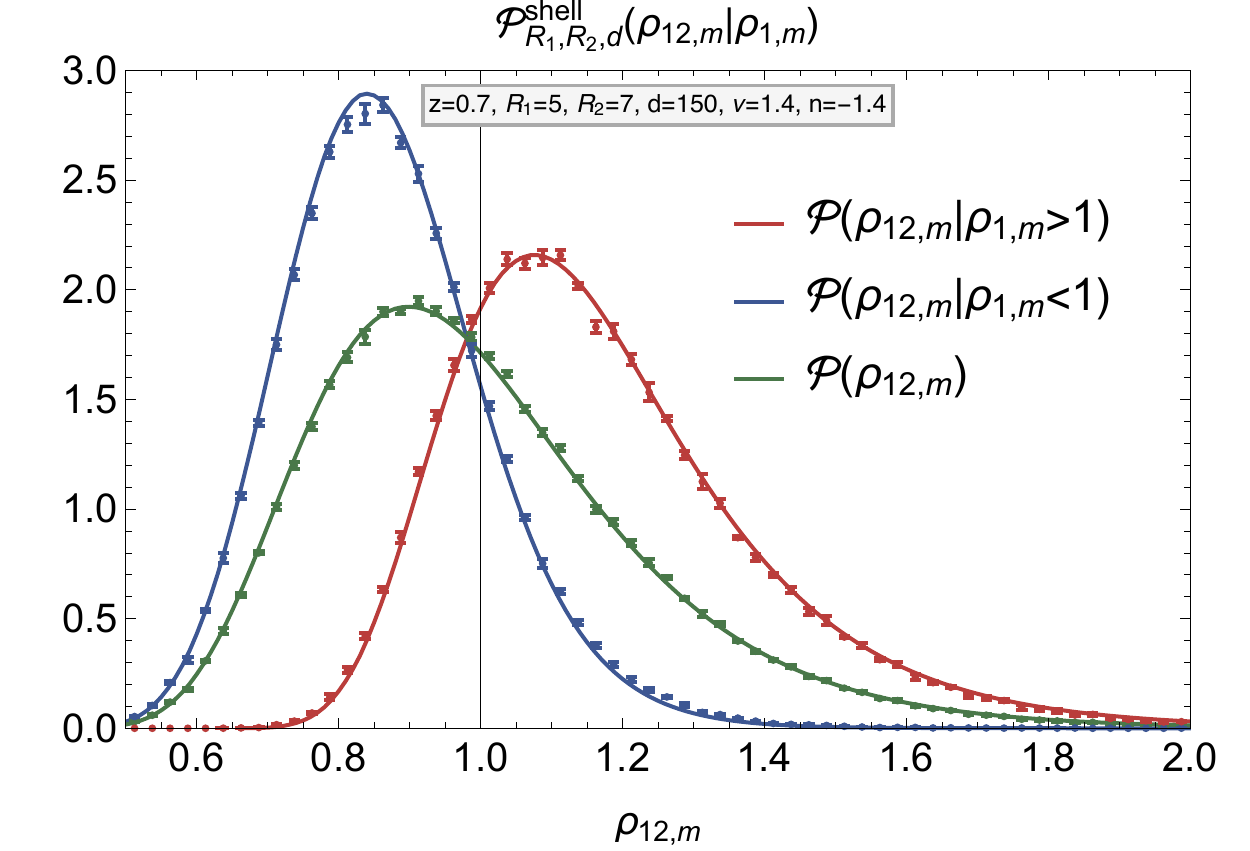}
\caption{Marginalised PDF of the matter densities in cylindrical rings of depth $d=150$ Mpc/$h$ and radii $R_1=5$ Mpc/$h$ and $R_2=7$ Mpc/$h$ at redshift $z=0.7$ from the saddle point approximation with $\nu=1.4$ and the disk filter for a power-law spectrum with $n=-1.4$ for arbitrary inner densities (green line), overdensities (red line) and underdensities (blue line) compared to the HR4 measurements (data points).}
   \label{fig:saddlePDFrho12condvsmeasurements}
\end{figure}

\begin{figure}
\includegraphics[width=\columnwidth]{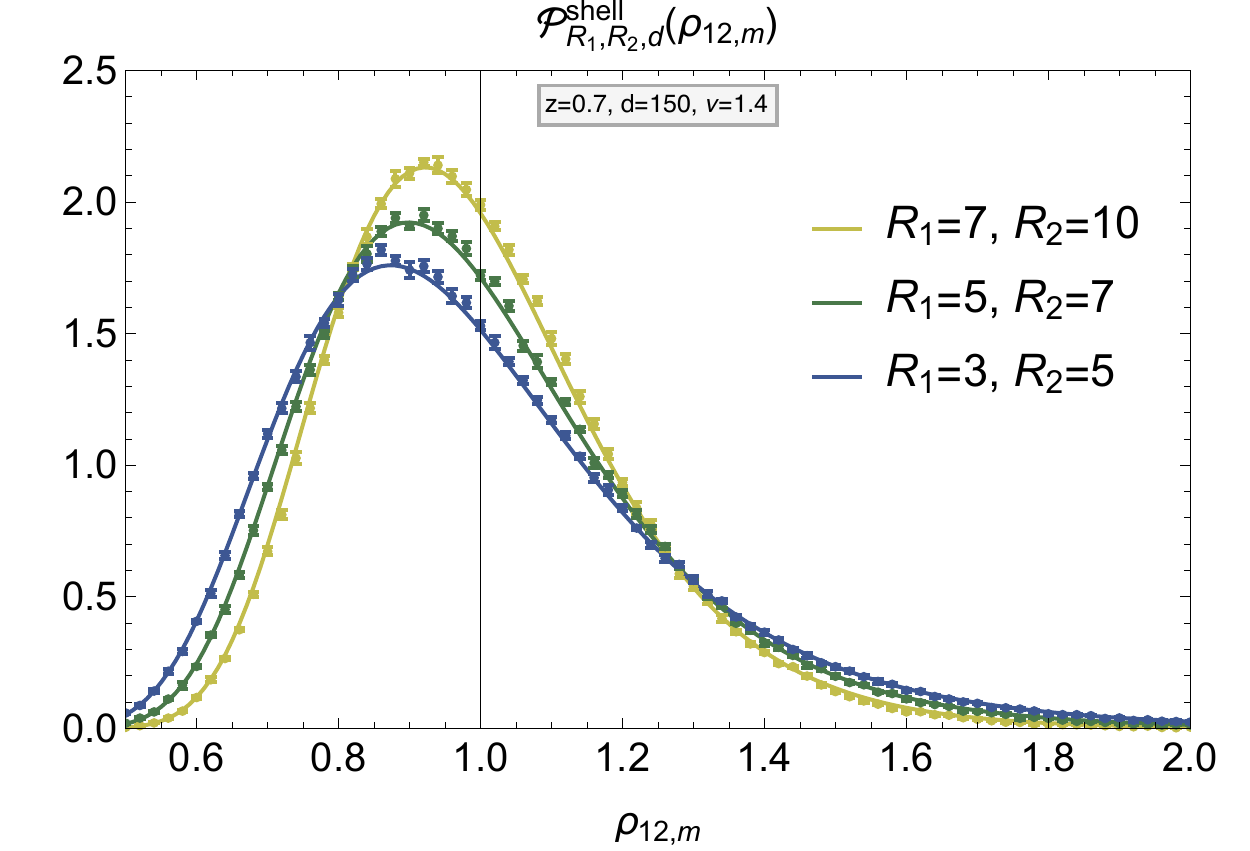}
\caption{Marginalised PDF of the matter densities in cylindrical shells of depth $d=150$ Mpc/$h$ and radii $(R_1,R_2)=\{(3,5),(5,7),(7,10)\}$ Mpc/$h$ at redshift $z=0.7$ from the saddle point approximation with $\nu=1.4$ and the disk filter for a power-law spectrum with $n=-\{1.5,1.4,1.3\}$ (colored lines) compared to the HR4 measurements (points with error bars).}
   \label{fig:saddlePDFrho12vsmeasurements}
\end{figure}

\subsection{Two-point cylinder bias}
In analogy to the 3D case leading to sphere bias described in \cite{Codis16correlations,Uhlemann17Kaiser}, cylinder bias encodes the density dependence of large-scale angular clustering. This idea is related to the concept of ‘sliced’ or ‘marked’ correlation functions, see for example early works of \cite{ShethTormen04, WhitePadmanabhan09} as well as \cite{Neyrinck16,White16}. Cylinder bias is defined as the ratio between the mean density $\rho'$ at separation $r_\perp$ (perpendicular to the axis of the cylinder) from a given density $\rho$ in a cylinder of radius $R$ and depth $d$ and the average angular-correlation function, see equation~\eqref{eq:defdensitybias}.
In the simulation, the cylinder bias is measured by using 4 neighbours at separation $r_\perp$ (along the two grid directions perpendicular to the cylinder axis) for every cylinder. A comparison between the prediction from equation~\eqref{eq:densitybias} using a cylindrical filter and the measurement from the simulation is shown in Fig.~\ref{fig:spherebias} for fixed cylinder depth and varying radius. Except for the very low density end, there is a very good agreement between the theoretical prediction and the measurement in the simulation. 

{The cylinder bias can be used to approximate the two-point PDF of densities in cylinders at large separation $r_\perp>2R$ according to
\begin{align}
 \mP_{\mR}^{\rm 2pt}(\rho_\rmm,\rho_\rmm',r_\perp) &= \mP_{\mR}^{\text{cyl}}(\rho_\rmm)\mP_{\mR}^{\text{cyl}}(\rho_\rmm')\\
\notag &\quad \times \left[1\!+\! \xi_{\rmm}(r_\perp) b_{\mR}^{\text{cyl}}(\rho_\rmm) b_\mR^{\text{cyl}}(\rho_\rmm') \right]\!, \label{eq:fullPDFlargeseparation}
\end{align}
and is hence also useful to quantify correlations between neighbouring cylinders that contribute to the error budget of PDFs \citep{Codis16correlations}.}

\begin{figure}
\includegraphics[width=\columnwidth]{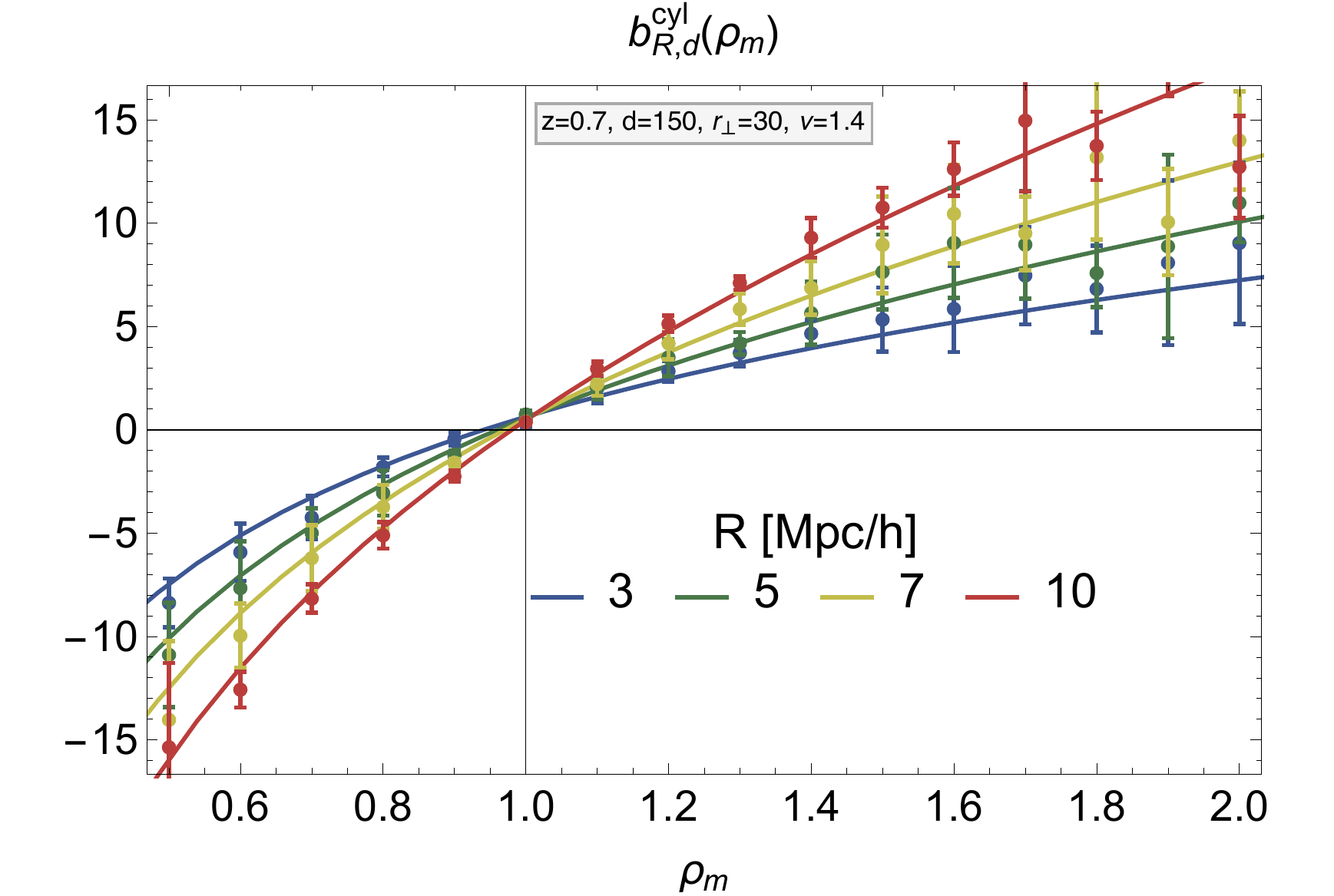}
\caption{Two-point cylinder bias of dark matter densities in cylinders of depth $d=150$ Mpc/$h$ and varying radii $R$ 
at separation $r_\perp=30$ Mpc$/h$ and redshift $z=0.7$ . The HR4 measurement (data points) is compared to the theoretical prediction from equation~\eqref{eq:densitybias} using the cylindrical filter.}
   \label{fig:spherebias}
\end{figure}
%
\section{Application to tracer densities} 
\label{sec:haloes}
  Let us now investigate   tracer counts using a  local bias relation. 
\subsection{Halo identification}
The haloes in HR4 were detected using Ordinary Parallel Friends-of-Friends
\citep[OPFOF,][]{KimPark06}, a massively parallel
implementation of the friends-of-friends 
(FoF) algorithm, using a canonical linking length of 0.2 mean particle
separations. Subhaloes were detected by the Physically Self-Bound algorithm
\citep[\textsc{psb},][]{KimPark06}, which finds the density peaks
within each FoF halo, removes  
unbound particles, similarly to the \textsc{subfind} halo finder, and additionally truncates the subhaloes to their
tidal radius.   All subhaloes with more than 30 particles were considered, yielding a masses from $2.7\times 10^{11}h ^{-1} M_\odot$ to $4.2 \times 10^{15}h ^{-1} M_\odot$.
Following the observations made in \cite{Seljak09,Hamaus10,Jee2012} and \cite{Uhlemann17Bias}, let us consider a mass-weighted halo density (instead of number-weighted) because this makes the bias relation much tighter and considerably reduces the scatter. This  can be intuitively understood as the mass-weighted halo densities resemble the overall dark matter density  more closely than halo number. Note however that the mass-weighted densities of subhaloes are expected to be very similar to the mass weighted density of  haloes (with no substructure) as the mass is almost preserved from haloes to subhaloes. Note finally that the subhaloes can be related to galaxies using for instance abundance matching techniques \citep{Kravtsov04,ValeOstriker04}. 

\subsection{A local bias model for halo densities in cylinders}
The treatment of mass-weighted halo densities in cylinders as biased tracers of the dark matter closely follows the procedure presented in \cite{Uhlemann17Bias} for densities in spheres. In order  to map the dark matter statistics to the halo statistics, let us rely on an `inverse' quadratic bias model $\mu_\rmm(\mu_\rmh)$ (writing the matter densities $\rho_{\rm m}$ as a function of the tracer densities $\rho_{\rm h}$) for the log-densities $\mu_{a}=\log\rho_{a}$ (a=m,h) which reads
\begin{equation}
\mu_{\rm m} =  \sum_{n=0}^{n_{\rm max}} b_{n}\mu_{\rmh}^{n}\ , \quad{\rm with} \quad\ n_{\rm max}=2\,.
\label{eq:POLYBIASloginv}
\end{equation}
Following the idea of \cite{Sigad2000,Szapudi2004}, a direct way to obtain the mean bias relation is to use the properties of the cumulative distribution functions (CDFs), defined as $\mC_{a}(\rho_{a})=\int_0^{\rho_{a}} d\rho' \mP_{a}^{\text{cyl}}(\rho')$ with index $a=\rmm$ for matter and $a=\rmh$ for halos,  so that
\begin{align}
\label{eq:CDFbias}
\mC_\rmm(\rho_\rmm)=\mC_\rmh(\rho_\rmh) \ \Rightarrow \ \rho_\rmm(\rho_\rmh)=\mC_\rmm^{-1}( \mC_\rmh(\rho_\rmh))\,.
\end{align}
This is used to verify the accuracy of the polynomial log-bias model by fitting the bias parameters in equation~\eqref{eq:POLYBIASloginv} to the parametrisation-independent bias function obtained from the CDFs and comparing the two resulting functions.

Fig.~\ref{fig:scatterplothalovsmatter} presents a scatter plot showing $\rho_{\rm h}$ as a function of $\rho_{\rm m}$ for redshift $z=0.7$, depth $d=150$ Mpc$/h$ and radius $R=5$ Mpc$/h$ in order to assess how well bias models characterise the halo density bias. The lines correspond to the mean bias obtained in a parametrisation-independent way from the CDF method (red line) and fits based on a quadratic bias model for the log-densities (dashed and solid orange line) according to equation~\eqref{eq:POLYBIASloginv}. The corresponding values of the best-fit bias parameters are given in Table~\ref{tab:biasfit} for different radii at $z=0.7$. The second-order bias model for the logarithmic densities based on equation~\eqref{eq:POLYBIASloginv} agrees almost perfectly with the parametrisation-independent way of inferring bias using CDFs as in equation~\eqref{eq:CDFbias} and matches simulation results very well. An approximation of the standard linear bias parameter between matter and halo densities can be obtained by expanding the logarithmic relation to get $\delta_h/\delta_m\simeq [\exp(b_0)b_1]^{-1}$ which gives values close to $2$ and slightly decreases with increasing radius.

\begin{table}
\centering
\begin{tabular}{|c|cc|cc|ccc|}
\hline
\multicolumn{1}{|c|}{} & \multicolumn{2}{c|}{variance} & \multicolumn{2}{c|}{correlation} &\multicolumn{3}{c|}{bias}  \\\hline
$R$ &  $\sigma_{\mu,\rmm} $& $\sigma_{\mu,\rmh} $& $\xi_{\rho,\rmm} $& $\xi_{\rho,\rmh} $ & $b_{0}$ & $ b_{1}$ & $ b_{2}$\\\hline
3 & 0.300 & 0.683 & 0.0088 & 0.0040 &  0.0199 & 0.4643 & 0.052\\
5 & 0.259 & 0.546 & 0.0081 & 0.0041 &  0.0184 & 0.4921 & 0.0562\\
7 & 0.232 & 0.472 & 0.0075 & 0.0042 &  0.0177 & 0.5067 & 0.0583\\
10 & 0.203 & 0.402 & 0.0067 & 0.0043 &  0.0175 & 0.5177 & 0.0627\\
\hline
\end{tabular}
\caption{Collection of simulation results for different radii $R$ and depth $d=150$ Mpc$/h$ at redshift $z=0.7$. The measured nonlinear variances $\sigma_\mu$ of the log-density $\mu=\log\rho$ and the correlation $\xi$ of the density $\rho$ at separation $r_\perp=30$Mpc$/h$ of both dark matter ($\rmm$) and haloes ($\rmh$) in real space 
along with the bias parameters obtained from fitting the quadratic model~\eqref{eq:POLYBIASloginv} to the CDF bias function~\eqref{eq:CDFbias}.}
\label{tab:biasfit}
\end{table} 

\begin{figure}
\center\includegraphics[width=0.75\columnwidth]{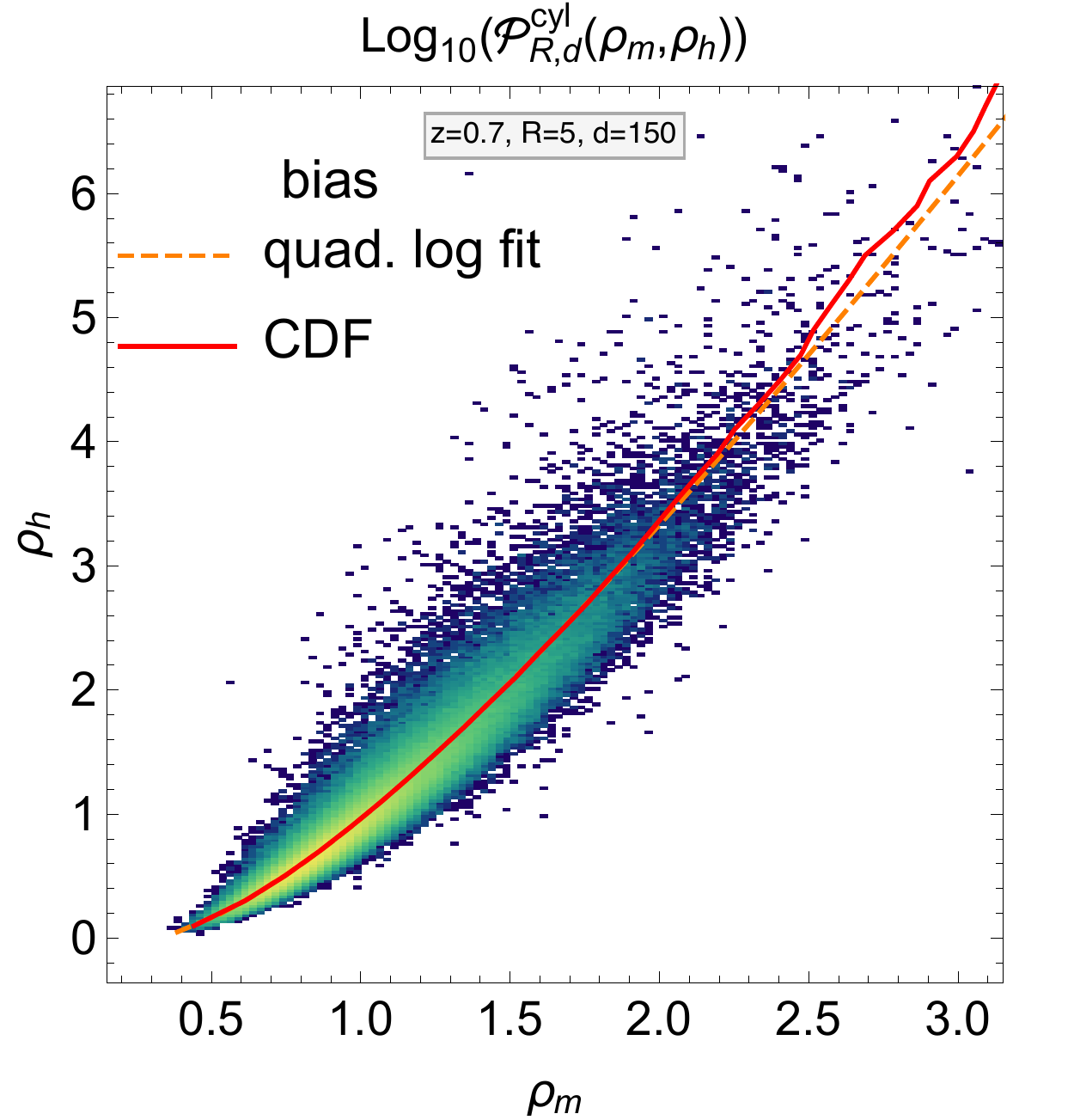}
\caption{Density scatter plot of the halo density $\rho_h$ with mass-weighting (blue-green region) versus the dark matter density $\rho_m$ for a cylinder of radius $R = 5$ Mpc$/h$ and depth $d=150$ Mpc$/h$ at redshift $z = 0.7$. The figure also shows the best-fit quadratic bias model for the log-density obtained from a fit to the scatter plot (dashed, orange) and to the CDF bias function (orange line, overlapping with the red line) which agrees very well with the parametrisation-independent bias obtained from the CDF (red line) in particular in the intermediate density region .
}
   \label{fig:scatterplothalovsmatter}
\end{figure}

\subsection{One-point PDF and two-point cylinder bias for tracers}
 Equipped with a bias model for the mean relation $\rho_\rmm(\rho_{\rmh})$, the halo PDF $\mP_{\rmh}$ is now obtained from the dark matter PDF $\mP_\rmm^{\text{cyl}}$ in equation~\eqref{eq:saddlePDFN-cell} by conservation of probability
\begin{equation}
\mP_{\rmh}^{\text{cyl}}\left (\rho_{\rmh}\right ) = \mP_\rmm^{\text{cyl}}(\rho_\rmm (\rho_{\rmh})) \left\lvert \dd\rho_\rmm/\dd\rho_{\rmh}\right\rvert\,.
\label{eq:HALOPDF}
\end{equation}
One can also obtain the modulation of the two-point correlation function, the cylinder bias $b_\circ$ for haloes, from the result for dark matter given in equation~\eqref{eq:densitybias}
\begin{equation}
\label{eq:spherebiashalo}
b_{\rmh}^{\text{cyl}}(\rho_\rmh) = b_{\rmm}^{\text{cyl}}\left(\rho_\rmm(\rho_\rmh)\right) \sqrt{\xi_{\rmm}/\xi_{\rmh}} \,,
\end{equation}
where the ratio of correlation functions is given by
\begin{equation}
\sqrt{\xi_{\rmh}/\xi_{\rmm}}=\left\langle \rho_{\rmh}(\rho_{\rmm}) b_{\rmm}^{\text{cyl}}(\rho_{\rmm})\right\rangle,
\end{equation} 
and can be approximated by expanding the log-bias relation to first order to obtain $\sqrt{\xi_{\rmm}/\xi_{\rmh}}
 \simeq \exp(b_0) b_1$. 
A validation of those theoretical predictions against the numerical simulation is shown in Fig.~\ref{fig:saddlePDFvsHorizoncylhalo} for the one point PDF and in Fig.~\ref{fig:spherebiashalo} for cylinder bias as function of halo density. The agreement is once again very good.
\begin{figure}
\includegraphics[width=\columnwidth]{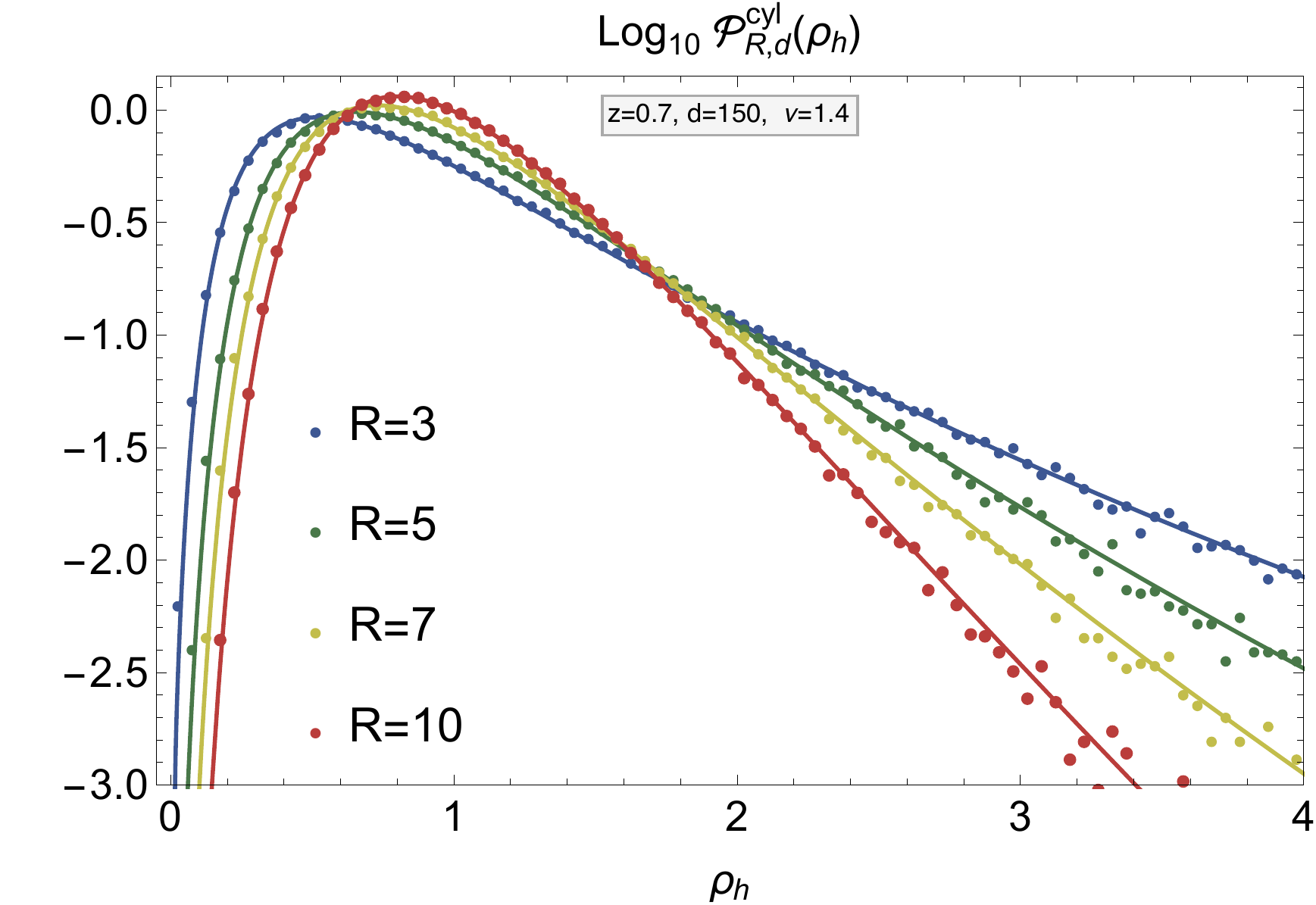}
\includegraphics[width=\columnwidth]{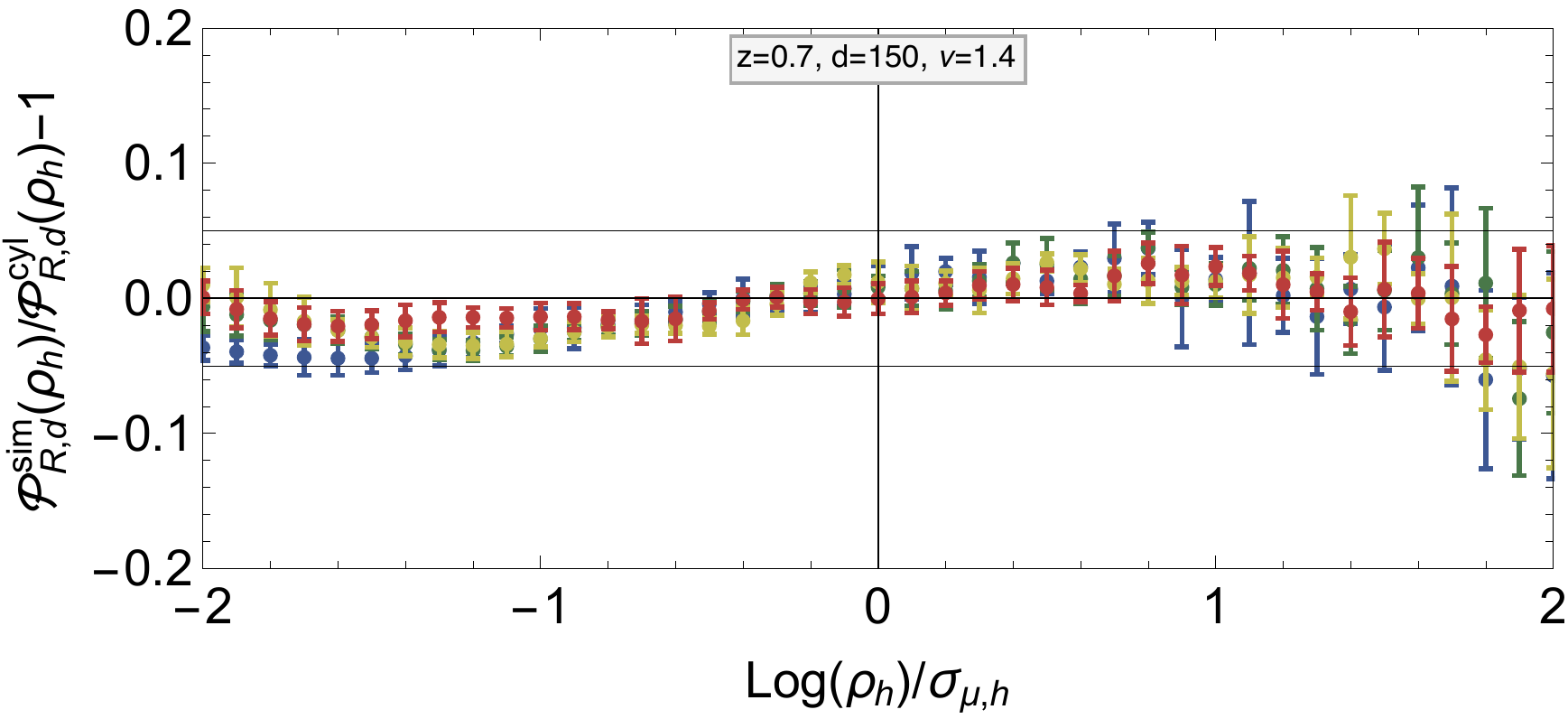}
\caption{PDF of the  mass-weighted halo density in cylindrical cells of length $d=150$ Mpc/$h$ and radii $R=3,5,7,10$ Mpc/$h$ at redshift $z=0.7$. Shown is a comparison of HR4 measurements (data points) and the prediction from equation~\eqref{eq:HALOPDF} computed with the cylindrical filter~\eqref{eq:filtercyl} together with the spherical collapse parametrized by $\nu=1.4$ (solid lines) and the fitted bias parameters from Table~\ref{tab:biasfit}. }
   \label{fig:saddlePDFvsHorizoncylhalo}
\end{figure}
\begin{figure}
\includegraphics[width=\columnwidth]{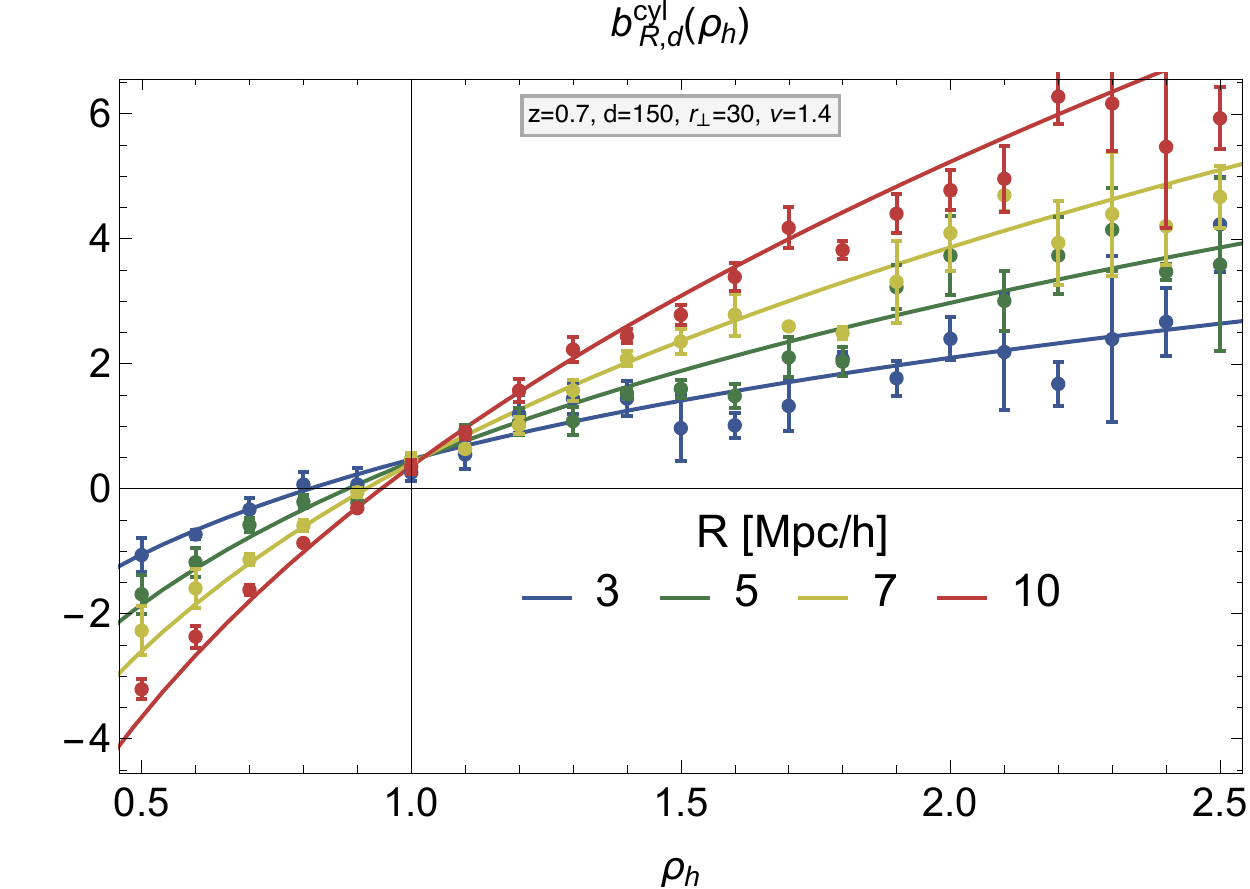}
\caption{Two-point cylinders bias of mass-weighted halo densities in cylinders of depth $d=150$ Mpc/$h$ and radii $R=3,5,7,10$ Mpc/$h$ at separation $r_\perp=30$ Mpc$/h$ and redshift $z=0.7$ . The HR4 measurement (data points) is compared to the theoretical prediction from equation~\eqref{eq:spherebiashalo} using the cylindrical filter and the best-fit bias parameters from Table~\ref{tab:biasfit}.}
   \label{fig:spherebiashalo}
\end{figure}
\subsection{Conditional PDF of DM density given tracer densities}
The conditional PDF of dark matter density in cylindrical rings given a central tracer density is interesting because it can be used to probe the matter density profile (for example measured through weak lensing) around regions of fixed tracer density. The matter density profile can be related to the tangential shear or lensing convergence profile around under- and overdense regions as is done in \cite{FriedrichDES17}. {The mixed halo matter joint PDF can be obtained from the joint PDF of dark matter using the one-to-one bias mapping
\begin{align}
\mP_{R_1\!,R_2}^{\text{cyl}}(\rho_{1,\rmh},\rho_{2,\rmm}) \!=\! \mP_{R_1\!,R_2}^{\text{cyl}}(\rho_{1,\rmm}(\rho_{1,\rmh}),\rho_{2,\rmm}) \left\lvert \frac{\dd\rho_{1,\rmm}}{\dd\rho_{1,\rmh}}\right\rvert\,,
\end{align}
and then defining conditionals   in analogy to equation~\eqref{eq:PDFshellcond}.} Similar to Figs.~\ref{fig:saddlePDF2cellvsmeasurements}~and~\ref{fig:saddlePDFrho12vsmeasurements} showing the joint PDF of matter in concentric cylinders and the conditional PDFs of density in rings given an inner dark matter over- or under-density, Fig.s~\ref{fig:saddlePDF2cellhmvsmeasurements}~and~\ref{fig:saddlePDFrho12hmvsmeasurements} show the case of a mixed joint PDF of dark matter around tracer density.

\subsection{Prospects for dark energy experiments} \label{sec:fiducial}
{Given the accuracy of the theoretical predictions for one and two-point statistics of halo densities in cylinders, counts-in-cells statistics seem promising for constraining cosmological parameters. In particular, measuring the tracer PDF in relatively thin redshift slices through photometric surveys gives access to the time evolution of clustering and hence can probe dark energy.}
Let us discuss briefly how the joint fit of estimators for counts-in-cells statistics presented in the previous  subsection 
could be used to constrain dark energy. 
In this context  the goal is to estimate the so-called equation of state of dark energy $w(z)=w_0+ w_a /({1+z})$ with parameters $(w_0,w_a)$ from the PDF while relying on
 the cosmic model for the growth rate \citep{Glazebrook}, 
\begin{align}
D(z|w_{0},w_a)&=\frac{5\Omega_m H_0^2}{2} H(a)\int_0^a \frac{{\rm d} a'}{a'^3 H^3(a')}\,, \\
H^2(a)=H_0^2 &\left[ \frac{\Omega_m}{a^3}+ \Omega_\Lambda \exp\left(3 \int_0^z \frac{1+w(z')}{1+z'} {\rm d} z'  \right)
 \right], \, \label{eq:cosmo}
\end{align}
 with $\Omega_m$, $\Omega_\Lambda$ and $H_0$ the dark matter and dark energy  densities and the Hubble constant respectively at redshift $z=0$ and $a\equiv1/(1+z)$ the expansion factor. {If one had access to the projected dark matter PDF at different redshift slices directly (through weak lensing convergence), one could follow the fiducial experiment described in \cite{Codis2016DE}. The key idea is to take advantage of the fact that the  variance is the only parameter in the dark matter PDF and that  its redshift-dependence is given by $\sigma(z)\propto D(z|w)$ where $D$ is given by equation~\eqref{eq:cosmo}.}
 
When applying this idea to galaxy counts, the main difficulty is that disentangling bias parameters and variance \citep[following][]{Uhlemann17Bias} {requires both one- and two-point statistics. Those are difficult to extract accurately enough on a thin slice-by-slice basis (such as that shown on Fig.~\ref{fig:HR4lightconesphere}), because the number of cylinders is limited by the available cosmic volume and the average number of galaxies per cylinders. In this case, one has to} circumvent this difficulty by parametrising the redshift evolution of the bias parameters (a possible parametrisation could be e.g. $b_n(z)=\sum a_{np}  z^p$). Then, one could marginalise over some prior values for this parametrisation {or use information at different scales (either varying radius or separation of the cylinders) to constrain those parameters jointly with $w$}. 
To constrain dark energy, one could naturally  compute the joint log-likelihood, ${\cal L}$ of the measured densities encoded in the halo  PDF as 
\begin{equation}\label{eq:like}
{\cal L}(\{(\rho_\rmh)_{i,k}\}|{w})=\sum_{k=1}^{N_z} \sum_{i=1}^{N_{\text{cyl}}(k)}\log {\cal P^{\text{cyl}}_{\mR}}\Big((\rho_{\rmh})_{i,k}|z_{k},{w}\Big)\,,
\end{equation}
where $N_z$ is the number of redshift slices, $N_{\text{cyl}}(k)$ the number of cylinders per slice and ${\cal P}\left((\rho_{\rmh})_{i,k}|z_k,{w}\right)$ is the theoretical density probability of having halo densities $(\rho_\rmh)_{i,k}$ in a cylinder of given radius at redshift-slice $z_k$ for a cosmological model with dark energy e.o.s parametrized by ${\it w}=(w_0,w_a)$ and  marginalized  over possible values of the bias parametrization. 
Optimizing ${\cal L}$ in equation~\eqref{eq:like} with respect to  $w$   
would yield a maximum likelihood estimate for the dark energy equation of state parameters.
As mentioned earlier, the key ingredient is to parametrize the redshift bias evolution with a few parameters,  so that each new redshift slice does not introduce new unknown parameters. {One could also attempt  taking correlations between neighbouring cylinders into account. The DE parameters could then be explored for example using an adapting (importance) sampling scheme where the sampling proposal could be a multi-variate Gaussian copula, see e.g. \cite{Benabed09Teasing}.} An alternative strategy to address the tracer bias problem would be to use a joint count and lensing-in-cell analysis as done in \cite{GruenDES17,FriedrichDES17}, possibly applied to tomographic bins.

Implementing the idea to constrain dark energy on the lightcone of HR4 will be the topic of future work
(but see Fig~\ref{fig:HR4lightconecounts}).

\begin{figure}
\includegraphics[width=\columnwidth]{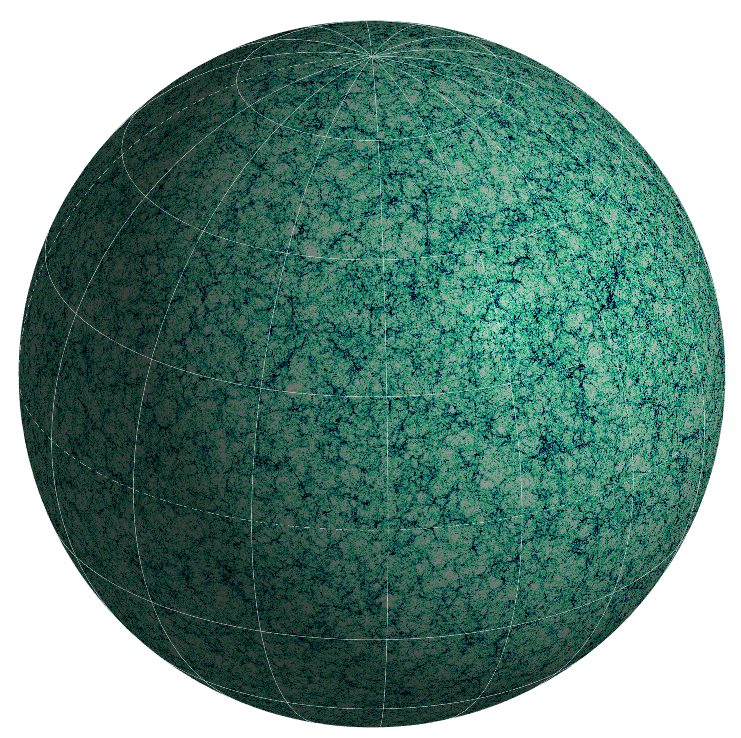}
\caption{Logarithmic view on the mass weighted count-in-cell of  sub-haloes in the HR4 lightcone 
at redshift {$z=0.36$} in a slice of 60 Mpc$/h$ projected on the sphere using the HealPix scheme (here $\ell_{\max}=512$). }
   \label{fig:HR4lightconesphere}
\end{figure}

\begin{figure}
\includegraphics[width=\columnwidth]{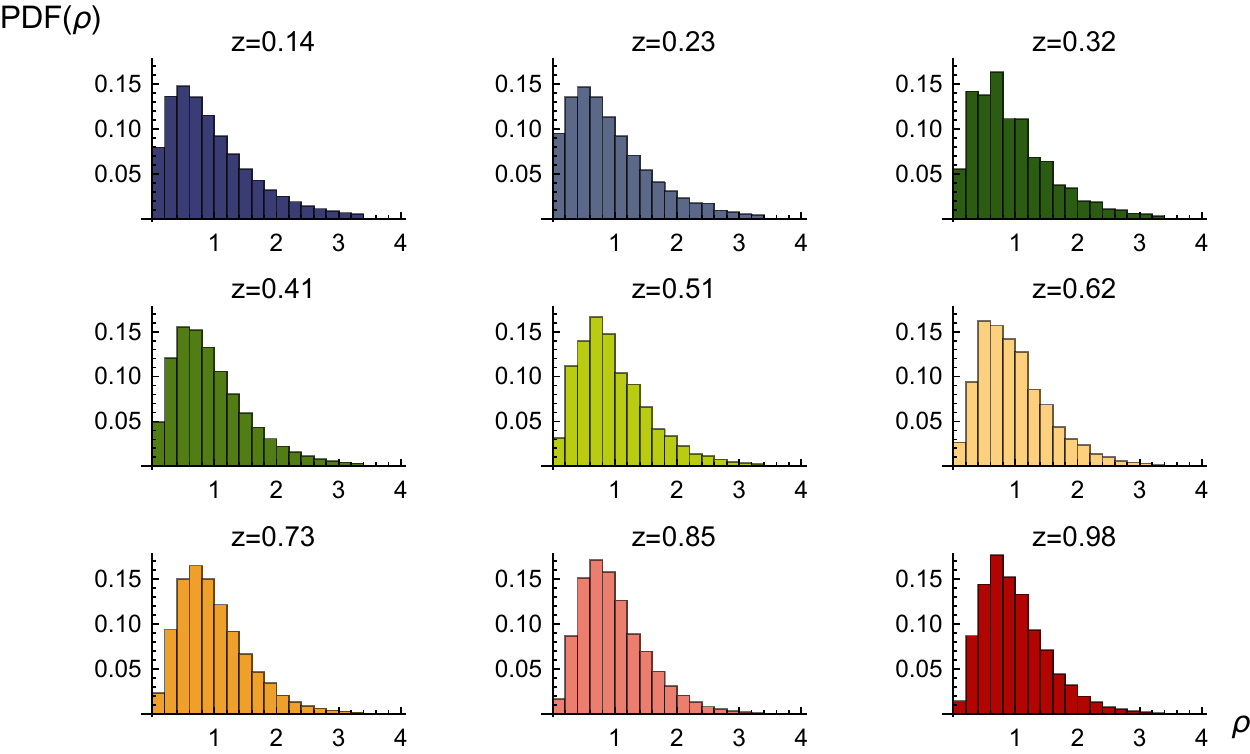}
\caption{Subhalo counts from the (non-cylindrical) HealPix cells for $\ell_{\max}=256$  also displayed  in Fig.~\ref{fig:HR4lightconesphere} at various redshifts as labeled. As expected, the PDFs become more Gaussian at higher redshift. Fitting counts like this following the procedure presented in Fig~\protect\ref{fig:saddlePDFvsHorizoncylhalo} 
should allow us to put constraints on the cosmic evolution of $\sigma$.}
   \label{fig:HR4lightconecounts}
\end{figure}

\section{Conclusion and Outlook}
\label{sec:conclusion}

Building upon recent works on large deviation statistics for cosmic densities in spherical cells,  estimators for  the one-point PDF of densities in concentric cylinders and for the density-dependence of their angular clustering (cylinder bias) -- therefore their two-point statistics -- were presented. 
These statistics were 
 combined with a simple polynomial  local bias model (in log-densities) relating dark matter and tracer densities in cylinders to produce 
 estimators for  2D count-in-cell statistics and compensated filters.
  The validity of the parametrised bias model was established using a parametrisation-independent extraction of the bias function.  
  All these estimators were found to be in excellent agreement with state-of-the-art numerical simulations. 
  The present formalism   will allow to probe cosmology using photometric surveys containing billions of galaxies. 
  It also provides the basis for a direct bias independent application to weak lensing mass aperture maps.
The PDF for dark matter and biased tracers could also be used to rapidly  generate  mock top-hat smoothed maps
for e.g. weak lensing or intensity mapping.

{A possible extension of this work is a validation of the analysis on lightcone slices with virtual galaxies extracted from Horizon Run 4. This could be used to quantify the accuracy at which dark energy parameters can be extracted from 2D count-in-cells following the fiducial dark energy experiment sketched in Sec.~\ref{sec:fiducial}. Another direction is to apply the large deviation statistics formalism to weak lensing convergence and shear in order to directly extract information from the dark matter density and its relation to biased tracers.}
The  PDF presented in equation~\eqref{eq:HALOPDF} could also be used 
\citep[while relying on a  fast particle-mesh code such as][to generate dark matter maps 
at the $\sim$ 5 Mpc$/h$ scale with accurate two-point functions]{Fengetal2016}    to rapidly  generate  mock top-hat smoothed  2D maps for dark haloes which would by construction have the right one and two-point statistics.

 Eventually, this formalism should be applied  to constrain cosmology  using  density and 2D counts-in-cells statistics  in ongoing or upcoming surveys like DES, Euclid, WFIRST, LSST, KiDs. 

\vskip 0.7cm
{\bf Acknowledgements:}  
 This work is partially supported by the grants ANR-12-BS05-0002 and  ANR-13-BS05-0005 of the French {\sl Agence Nationale de la Recherche}. CU was supported by the Delta-ITP consortium, a programme of the Netherlands organization for scientific research (NWO) that is funded by the Dutch Ministry of Education, Culture and Science (OCW). CU also kindly acknowledges funding by the STFC grant RG84196 `Revealing the Structure of the Universe'. 
  We thank the KIAS Center for Advanced Computation for providing computing resources. CU thanks IAP, CITA, KIAS and KASI for hospitality during the initiation, evolution and completion of this project. SP thanks KIAS for their hospitality.
 Special thanks to Steven Appleby for his help with Fig~\ref{fig:HR4lightconesphere}.
We acknowledge support from S. Rouberol for running the horizon cluster hosted by the Institut d'Astrophysique de Paris.


\bibliographystyle{mnras}
\bibliography{LSStructure}

\begin{thebibliography}{}
\makeatletter
\relax
\def\mn@urlcharsother{\let\do\@makeother \do\$\do\&\do\#\do\^\do\_\do\%\do\~}
\def\mn@doi{\begingroup\mn@urlcharsother \@ifnextchar [ {\mn@doi@}
  {\mn@doi@[]}}
\def\mn@doi@[#1]#2{\def\@tempa{#1}\ifx\@tempa\@empty \href
  {http://dx.doi.org/#2} {doi:#2}\else \href {http://dx.doi.org/#2} {#1}\fi
  \endgroup}
\def\mn@eprint#1#2{\mn@eprint@#1:#2::\@nil}
\def\mn@eprint@arXiv#1{\href {http://arxiv.org/abs/#1} {{\tt arXiv:#1}}}
\def\mn@eprint@dblp#1{\href {http://dblp.uni-trier.de/rec/bibtex/#1.xml}
  {dblp:#1}}
\def\mn@eprint@#1:#2:#3:#4\@nil{\def\@tempa {#1}\def\@tempb {#2}\def\@tempc
  {#3}\ifx \@tempc \@empty \let \@tempc \@tempb \let \@tempb \@tempa \fi \ifx
  \@tempb \@empty \def\@tempb {arXiv}\fi \@ifundefined
  {mn@eprint@\@tempb}{\@tempb:\@tempc}{\expandafter \expandafter \csname
  mn@eprint@\@tempb\endcsname \expandafter{\@tempc}}}

\bibitem[\protect\citeauthoryear{{Agrawal}, {Makiya}, {Chiang}, {Jeong},
  {Saito}  \& {Komatsu}}{{Agrawal} et~al.}{2017}]{Agrawal17}
{Agrawal} A.,  {Makiya} R.,  {Chiang} C.-T.,  {Jeong} D.,  {Saito} S.,
  {Komatsu} E.,  2017, \mn@doi [\jcap] {10.1088/1475-7516/2017/10/003}, \href
  {http://adsabs.harvard.edu/abs/2017JCAP...10..003A} {10, 003}

\bibitem[\protect\citeauthoryear{{Bel} \& {Marinoni}}{{Bel} \&
  {Marinoni}}{2014}]{BelMarinoni14}
{Bel} J.,  {Marinoni} C.,  2014, \mn@doi [\aap] {10.1051/0004-6361/201321941},
  \href {http://adsabs.harvard.edu/abs/2014A%26A...563A..36B} {563, A36}

\bibitem[\protect\citeauthoryear{{Bel} et~al.,}{{Bel} et~al.}{2016}]{Bel16}
{Bel} J.,  et~al., 2016, \mn@doi [\aap] {10.1051/0004-6361/201526455}, \href
  {http://adsabs.harvard.edu/abs/2016A%26A...588A..51B} {588, A51}

\bibitem[\protect\citeauthoryear{{Benabed}, {Cardoso}, {Prunet}  \&
  {Hivon}}{{Benabed} et~al.}{2009}]{Benabed09Teasing}
{Benabed} K.,  {Cardoso} J.-F.,  {Prunet} S.,   {Hivon} E.,  2009, \mn@doi
  [\mnras] {10.1111/j.1365-2966.2009.15202.x}, \href
  {http://adsabs.harvard.edu/abs/2009MNRAS.400..219B} {400, 219}

\bibitem[\protect\citeauthoryear{{Bernardeau}}{{Bernardeau}}{1992}]{Bernardeau92}
{Bernardeau} F.,  1992, \mn@doi [\apj] {10.1086/171398}, \href
  {http://adsabs.harvard.edu/cgi-bin/nph-bib_query?bibcode=1992ApJ...392....1B&db_key=AST}
  {392, 1}

\bibitem[\protect\citeauthoryear{{Bernardeau}}{{Bernardeau}}{1994}]{Bernardeau94smooth}
{Bernardeau} F.,  1994, \aap, \href
  {http://adsabs.harvard.edu/cgi-bin/nph-bib_query?bibcode=1994A%26A...291..697B&db_key=AST}
  {291, 697}

\bibitem[\protect\citeauthoryear{{Bernardeau}}{{Bernardeau}}{1995}]{Bernardeau1995}
{Bernardeau} F.,  1995, \aap, \href
  {http://adsabs.harvard.edu/cgi-bin/nph-bib_query?bibcode=1995A%26A...301..309B&db_key=AST}
  {301, 309}

\bibitem[\protect\citeauthoryear{{Bernardeau}}{{Bernardeau}}{1996}]{Bernardeau96bias}
{Bernardeau} F.,  1996, \aap, \href
  {http://adsabs.harvard.edu/cgi-bin/nph-bib_query?bibcode=1996A%26A...312...11B&db_key=AST}
  {312, 11}

\bibitem[\protect\citeauthoryear{{Bernardeau} \& {Reimberg}}{{Bernardeau} \&
  {Reimberg}}{2016}]{BernardeauReimberg15}
{Bernardeau} F.,  {Reimberg} P.,  2016, \mn@doi [\prd]
  {10.1103/PhysRevD.94.063520}, \href
  {http://adsabs.harvard.edu/abs/2015arXiv151108641B} {94, 063520}

\bibitem[\protect\citeauthoryear{{Bernardeau} \& {Valageas}}{{Bernardeau} \&
  {Valageas}}{2000}]{BV2000}
{Bernardeau} F.,  {Valageas} P.,  2000, \aap, \href
  {http://adsabs.harvard.edu/abs/2000A%26A...364....1B} {364, 1}

\bibitem[\protect\citeauthoryear{{Bernardeau}, {Colombi}, {Gazta{\~n}aga}  \&
  {Scoccimarro}}{{Bernardeau} et~al.}{2002}]{Bernardeau02LSS}
{Bernardeau} F.,  {Colombi} S.,  {Gazta{\~n}aga} E.,   {Scoccimarro} R.,  2002,
  \physrep, \href
  {http://adsabs.harvard.edu/cgi-bin/nph-bib_query?bibcode=2002PhR...367....1B&db_key=AST}
  {367, 1}

\bibitem[\protect\citeauthoryear{{Bernardeau}, {Pichon}  \&
  {Codis}}{{Bernardeau} et~al.}{2014a}]{Bernardeau14}
{Bernardeau} F.,  {Pichon} C.,   {Codis} S.,  2014a, \mn@doi [\prd]
  {10.1103/PhysRevD.90.103519}, \href
  {http://adsabs.harvard.edu/abs/2014PhRvD..90j3519B} {90, 103519}

\bibitem[\protect\citeauthoryear{{Bernardeau}, {Pichon}  \&
  {Codis}}{{Bernardeau} et~al.}{2014b}]{Bernardeau2014}
{Bernardeau} F.,  {Pichon} C.,   {Codis} S.,  2014b, \mn@doi [\prd]
  {10.1103/PhysRevD.90.103519}, \href
  {http://adsabs.harvard.edu/abs/2014PhRvD..90j3519B} {90, 103519}

\bibitem[\protect\citeauthoryear{{Blake} \& {Glazebrook}}{{Blake} \&
  {Glazebrook}}{2003}]{Glazebrook}
{Blake} C.,  {Glazebrook} K.,  2003, \mn@doi [\apj] {10.1086/376983}, \href
  {http://cdsads.u-strasbg.fr/abs/2003ApJ...594..665B} {594, 665}

\bibitem[\protect\citeauthoryear{{Bouchet}, {Strauss}, {Davis}, {Fisher},
  {Yahil}  \& {Huchra}}{{Bouchet} et~al.}{1993}]{1993ApJ...417...36B}
{Bouchet} F.~R.,  {Strauss} M.~A.,  {Davis} M.,  {Fisher} K.~B.,  {Yahil} A.,
  {Huchra} J.~P.,  1993, \mn@doi [\apj] {10.1086/173289}, \href
  {http://cdsads.u-strasbg.fr/abs/1993ApJ...417...36B} {417, 36}

\bibitem[\protect\citeauthoryear{{Clerkin} et~al.,}{{Clerkin}
  et~al.}{2017}]{Clerkin17}
{Clerkin} L.,  et~al., 2017, \mn@doi [\mnras] {10.1093/mnras/stw2106}, \href
  {http://adsabs.harvard.edu/abs/2017MNRAS.466.1444C} {466, 1444}

\bibitem[\protect\citeauthoryear{{Codis}, {Pichon}, {Bernardeau}, {Uhlemann}
  \& {Prunet}}{{Codis} et~al.}{2016a}]{Codis2016DE}
{Codis} S.,  {Pichon} C.,  {Bernardeau} F.,  {Uhlemann} C.,   {Prunet} S.,
  2016a, \mn@doi [\mnras] {10.1093/mnras/stw1084}, \href
  {http://adsabs.harvard.edu/abs/2016MNRAS.460.1549C} {460, 1549}

\bibitem[\protect\citeauthoryear{{Codis}, {Bernardeau}  \& {Pichon}}{{Codis}
  et~al.}{2016b}]{Codis16correlations}
{Codis} S.,  {Bernardeau} F.,   {Pichon} C.,  2016b, \mn@doi [\mnras]
  {10.1093/mnras/stw1103}, \href
  {http://cdsads.u-strasbg.fr/abs/2016MNRAS.460.1598C} {460, 1598}

\bibitem[\protect\citeauthoryear{{Coles} \& {Jones}}{{Coles} \&
  {Jones}}{1991}]{ColesJones91}
{Coles} P.,  {Jones} B.,  1991, \mn@doi [\mnras] {10.1093/mnras/248.1.1}, \href
  {http://adsabs.harvard.edu/abs/1991MNRAS.248....1C} {248, 1}

\bibitem[\protect\citeauthoryear{{DESI Collaboration} et~al.,}{{DESI
  Collaboration} et~al.}{2016}]{DESI}
{DESI Collaboration} et~al., 2016, preprint, \href
  {http://adsabs.harvard.edu/abs/2016arXiv161100036D} {} (\mn@eprint {arXiv}
  {1611.00036})

\bibitem[\protect\citeauthoryear{{Driver} et~al.,}{{Driver}
  et~al.}{2016}]{Driver2016}
{Driver} S.~P.,  et~al., 2016, \mn@doi [\mnras] {10.1093/mnras/stv2505}, \href
  {http://adsabs.harvard.edu/abs/2016MNRAS.455.3911D} {455, 3911}

\bibitem[\protect\citeauthoryear{{Dubinski}, {Kim}, {Park}  \&
  {Humble}}{{Dubinski} et~al.}{2004}]{GOTPM}
{Dubinski} J.,  {Kim} J.,  {Park} C.,   {Humble} R.,  2004, \mn@doi [\na]
  {10.1016/j.newast.2003.08.002}, \href
  {http://adsabs.harvard.edu/abs/2004NewA....9..111D} {9, 111}

\bibitem[\protect\citeauthoryear{{Efstathiou}}{{Efstathiou}}{1995}]{Efstathiou95}
{Efstathiou} G.,  1995, \mn@doi [\mnras] {10.1093/mnras/276.4.1425}, \href
  {http://adsabs.harvard.edu/abs/1995MNRAS.276.1425E} {276, 1425}

\bibitem[\protect\citeauthoryear{{Feng}, {Chu}, {Seljak}  \& {McDonald}}{{Feng}
  et~al.}{2016}]{Fengetal2016}
{Feng} Y.,  {Chu} M.-Y.,  {Seljak} U.,   {McDonald} P.,  2016, \mn@doi [\mnras]
  {10.1093/mnras/stw2123}, \href
  {http://adsabs.harvard.edu/abs/2016MNRAS.463.2273F} {463, 2273}

\bibitem[\protect\citeauthoryear{{Friedrich} et~al.,}{{Friedrich}
  et~al.}{2017}]{FriedrichDES17}
{Friedrich} O.,  et~al., 2017, preprint, \href
  {http://adsabs.harvard.edu/abs/2017arXiv171005162F} {} (\mn@eprint {arXiv}
  {1710.05162})

\bibitem[\protect\citeauthoryear{{Gruen} et~al.,}{{Gruen}
  et~al.}{2016}]{Gruen16troughs}
{Gruen} D.,  et~al., 2016, \mn@doi [\mnras] {10.1093/mnras/stv2506}, \href
  {http://adsabs.harvard.edu/abs/2016MNRAS.455.3367G} {455, 3367}

\bibitem[\protect\citeauthoryear{{Gruen} et~al.,}{{Gruen}
  et~al.}{2017}]{GruenDES17}
{Gruen} D.,  et~al., 2017, preprint, \href
  {http://adsabs.harvard.edu/abs/2017arXiv171005045G} {} (\mn@eprint {arXiv}
  {1710.05045})

\bibitem[\protect\citeauthoryear{{Guzzo} et~al.,}{{Guzzo} et~al.}{2014}]{VIPER}
{Guzzo} L.,  et~al., 2014, \mn@doi [\aap] {10.1051/0004-6361/201321489}, \href
  {http://adsabs.harvard.edu/abs/2014A%26A...566A.108G} {566, A108}

\bibitem[\protect\citeauthoryear{{Hamaus}, {Seljak}, {Desjacques}, {Smith}  \&
  {Baldauf}}{{Hamaus} et~al.}{2010}]{Hamaus10}
{Hamaus} N.,  {Seljak} U.,  {Desjacques} V.,  {Smith} R.~E.,   {Baldauf} T.,
  2010, \mn@doi [\prd] {10.1103/PhysRevD.82.043515}, \href
  {http://adsabs.harvard.edu/abs/2010PhRvD..82d3515H} {82, 043515}

\bibitem[\protect\citeauthoryear{{Hilbert}, {Hartlap}  \&
  {Schneider}}{{Hilbert} et~al.}{2011}]{Hilbert11}
{Hilbert} S.,  {Hartlap} J.,   {Schneider} P.,  2011, \mn@doi [\aap]
  {10.1051/0004-6361/201117294}, \href
  {http://adsabs.harvard.edu/abs/2011A%26A...536A..85H} {536, A85}

\bibitem[\protect\citeauthoryear{{Hurtado-Gil}, {Mart{\'{\i}}nez},
  {Arnalte-Mur}, {Pons-Border{\'{\i}}a}, {Pareja-Flores}  \&
  {Paredes}}{{Hurtado-Gil} et~al.}{2017}]{Hurtado-Gil17}
{Hurtado-Gil} L.,  {Mart{\'{\i}}nez} V.~J.,  {Arnalte-Mur} P.,
  {Pons-Border{\'{\i}}a} M.-J.,  {Pareja-Flores} C.,   {Paredes} S.,  2017,
  \mn@doi [\aap] {10.1051/0004-6361/201629097}, \href
  {http://adsabs.harvard.edu/abs/2017A%26A...601A..40H} {601, A40}

\bibitem[\protect\citeauthoryear{{Jee}, {Park}, {Kim}, {Choi}  \& {Kim}}{{Jee}
  et~al.}{2012}]{Jee2012}
{Jee} I.,  {Park} C.,  {Kim} J.,  {Choi} Y.-Y.,   {Kim} S.~S.,  2012, \mn@doi
  [\apj] {10.1088/0004-637X/753/1/11}, \href
  {http://adsabs.harvard.edu/abs/2012ApJ...753...11J} {753, 11}

\bibitem[\protect\citeauthoryear{{Kacprzak}}{{Kacprzak}}{2016}]{KacprzakDES16}
{Kacprzak} T. e.~a.,  2016, preprint, \href
  {http://adsabs.harvard.edu/abs/2016arXiv160305040K} {} (\mn@eprint {arXiv}
  {1603.05040})

\bibitem[\protect\citeauthoryear{{Kaiser}}{{Kaiser}}{1984}]{Kaiser84}
{Kaiser} N.,  1984, \mn@doi [\apjl] {10.1086/184341}, \href
  {http://adsabs.harvard.edu/abs/1984ApJ...284L...9K} {284, L9}

\bibitem[\protect\citeauthoryear{{Kayo}, {Taruya}  \& {Suto}}{{Kayo}
  et~al.}{2001}]{Kayo01}
{Kayo} I.,  {Taruya} A.,   {Suto} Y.,  2001, \mn@doi [\apj] {10.1086/323227},
  \href {http://adsabs.harvard.edu/abs/2001ApJ...561...22K} {561, 22}

\bibitem[\protect\citeauthoryear{{Kim} \& {Park}}{{Kim} \&
  {Park}}{2006}]{KimPark06}
{Kim} J.,  {Park} C.,  2006, \mn@doi [\apj] {10.1086/499761}, \href
  {http://adsabs.harvard.edu/abs/2006ApJ...639..600K} {639, 600}

\bibitem[\protect\citeauthoryear{{Kim}, {Park}, {L'Huillier}  \& {Hong}}{{Kim}
  et~al.}{2015}]{HR4}
{Kim} J.,  {Park} C.,  {L'Huillier} B.,   {Hong} S.~E.,  2015, \mn@doi [J.
  Korean Astron. Soc.] {10.5303/JKAS.2015.48.4.213}, \href
  {http://adsabs.harvard.edu/abs/2015JKAS...48..213K} {48, 213}

\bibitem[\protect\citeauthoryear{{Kravtsov}, {Berlind}, {Wechsler}, {Klypin},
  {Gottl{\"o}ber}, {Allgood}  \& {Primack}}{{Kravtsov}
  et~al.}{2004}]{Kravtsov04}
{Kravtsov} A.~V.,  {Berlind} A.~A.,  {Wechsler} R.~H.,  {Klypin} A.~A.,
  {Gottl{\"o}ber} S.,  {Allgood} B.,   {Primack} J.~R.,  2004, \mn@doi [\apj]
  {10.1086/420959}, \href {http://adsabs.harvard.edu/abs/2004ApJ...609...35K}
  {609, 35}

\bibitem[\protect\citeauthoryear{{L'Huillier}, {Park}  \& {Kim}}{{L'Huillier}
  et~al.}{2014}]{LHuillier14}
{L'Huillier} B.,  {Park} C.,   {Kim} J.,  2014, \mn@doi [\na]
  {10.1016/j.newast.2014.01.007}, \href
  {http://adsabs.harvard.edu/abs/2014NewA...30...79L} {30, 79}

\bibitem[\protect\citeauthoryear{{LSST Dark Energy Science
  Collaboration}}{{LSST Dark Energy Science Collaboration}}{2012}]{LSST}
{LSST Dark Energy Science Collaboration} 2012, preprint, \href
  {http://adsabs.harvard.edu/abs/2012arXiv1211.0310L} {} (\mn@eprint {arXiv}
  {1211.0310})

\bibitem[\protect\citeauthoryear{{Laureijs} et~al.,}{{Laureijs}
  et~al.}{2011}]{Euclid}
{Laureijs} R.,  et~al., 2011, preprint, \href
  {http://adsabs.harvard.edu/abs/2011arXiv1110.3193L} {} (\mn@eprint {arXiv}
  {1110.3193})

\bibitem[\protect\citeauthoryear{{McConnachie}, {Babusiaux}, {Balogh},
  {Driver}, {C{\^o}t{\'e}}  \& et al.}{{McConnachie}
  et~al.}{2016}]{2016arXiv160600043M}
{McConnachie} A.,  {Babusiaux} C.,  {Balogh} M.,  {Driver} S.,  {C{\^o}t{\'e}}
  et al. 2016, preprint, \href
  {http://cdsads.u-strasbg.fr/abs/2016arXiv160600043M} {} (\mn@eprint {arXiv}
  {1606.00043})

\bibitem[\protect\citeauthoryear{{Munshi} \& {Jain}}{{Munshi} \&
  {Jain}}{2000}]{Munshi00}
{Munshi} D.,  {Jain} B.,  2000, \mn@doi [\mnras]
  {10.1046/j.1365-8711.2000.03694.x}, \href
  {http://adsabs.harvard.edu/abs/2000MNRAS.318..109M} {318, 109}

\bibitem[\protect\citeauthoryear{{Munshi} \& {Jain}}{{Munshi} \&
  {Jain}}{2001}]{Munshi01}
{Munshi} D.,  {Jain} B.,  2001, \mn@doi [\mnras]
  {10.1046/j.1365-8711.2001.04069.x}, \href
  {http://adsabs.harvard.edu/abs/2001MNRAS.322..107M} {322, 107}

\bibitem[\protect\citeauthoryear{{Munshi}, {Valageas}  \& {Barber}}{{Munshi}
  et~al.}{2004}]{Munshi04}
{Munshi} D.,  {Valageas} P.,   {Barber} A.~J.,  2004, \mn@doi [\mnras]
  {10.1111/j.1365-2966.2004.07553.x}, \href
  {http://adsabs.harvard.edu/abs/2004MNRAS.350...77M} {350, 77}

\bibitem[\protect\citeauthoryear{{Munshi}, {Coles}  \& {Kilbinger}}{{Munshi}
  et~al.}{2014}]{Munshi14}
{Munshi} D.,  {Coles} P.,   {Kilbinger} M.,  2014, \mn@doi [\jcap]
  {10.1088/1475-7516/2014/04/004}, \href
  {http://adsabs.harvard.edu/abs/2014JCAP...04..004M} {4, 004}

\bibitem[\protect\citeauthoryear{{Neyrinck}, {Szapudi}, {McCullagh}, {Szalay},
  {Falck}  \& {Wang}}{{Neyrinck} et~al.}{2016}]{Neyrinck16}
{Neyrinck} M.~C.,  {Szapudi} I.,  {McCullagh} N.,  {Szalay} A.,  {Falck} B.,
  {Wang} J.,  2016, preprint, \href
  {http://adsabs.harvard.edu/abs/2016arXiv161006215N} {} (\mn@eprint {arXiv}
  {1610.06215})

\bibitem[\protect\citeauthoryear{{Pajer} \& {van der Woude}}{{Pajer} \& {van
  der Woude}}{2017}]{vanderWoude017}
{Pajer} E.,  {van der Woude} D.,  2017, preprint, \href
  {http://adsabs.harvard.edu/abs/2017arXiv171001736P} {} (\mn@eprint {arXiv}
  {1710.01736})

\bibitem[\protect\citeauthoryear{{Reimberg} \& {Bernardeau}}{{Reimberg} \&
  {Bernardeau}}{2017}]{ReimbergBernardeau17}
{Reimberg} P.,  {Bernardeau} F.,  2017, preprint, \href
  {http://adsabs.harvard.edu/abs/2017arXiv170800252R} {} (\mn@eprint {arXiv}
  {1708.00252})

\bibitem[\protect\citeauthoryear{{Rykoff} et~al.,}{{Rykoff}
  et~al.}{2016}]{DES2016}
{Rykoff} E.~S.,  et~al., 2016, \mn@doi [\apjs] {10.3847/0067-0049/224/1/1},
  \href {http://adsabs.harvard.edu/abs/2016ApJS..224....1R} {224, 1}

\bibitem[\protect\citeauthoryear{{Seljak}}{{Seljak}}{2009}]{Seljak09}
{Seljak} U.,  2009, \mn@doi [Physical Review Letters]
  {10.1103/PhysRevLett.102.021302}, \href
  {http://adsabs.harvard.edu/abs/2009PhRvL.102b1302S} {102, 021302}

\bibitem[\protect\citeauthoryear{{Sheth} \& {Tormen}}{{Sheth} \&
  {Tormen}}{2004}]{ShethTormen04}
{Sheth} R.~K.,  {Tormen} G.,  2004, \mn@doi [\mnras]
  {10.1111/j.1365-2966.2004.07733.x}, \href
  {http://adsabs.harvard.edu/abs/2004MNRAS.350.1385S} {350, 1385}

\bibitem[\protect\citeauthoryear{{Sigad}, {Branchini}  \& {Dekel}}{{Sigad}
  et~al.}{2000}]{Sigad2000}
{Sigad} Y.,  {Branchini} E.,   {Dekel} A.,  2000, \mn@doi [\apj]
  {10.1086/309331}, \href {http://adsabs.harvard.edu/abs/2000ApJ...540...62S}
  {540, 62}

\bibitem[\protect\citeauthoryear{{Spergel} et~al.,}{{Spergel}
  et~al.}{2013}]{WFIRST}
{Spergel} D.,  et~al., 2013, preprint, \href
  {http://cdsads.u-strasbg.fr/abs/2013arXiv1305.5422S} {} (\mn@eprint {arXiv}
  {1305.5422})

\bibitem[\protect\citeauthoryear{{Szapudi} \& {Pan}}{{Szapudi} \&
  {Pan}}{2004}]{Szapudi2004}
{Szapudi} I.,  {Pan} J.,  2004, \mn@doi [\apj] {10.1086/380920}, \href
  {http://adsabs.harvard.edu/abs/2004ApJ...602...26S} {602, 26}

\bibitem[\protect\citeauthoryear{{Takada} et~al.,}{{Takada} et~al.}{2014}]{PFS}
{Takada} M.,  et~al., 2014, \mn@doi [\pasj] {10.1093/pasj/pst019}, \href
  {http://adsabs.harvard.edu/abs/2014PASJ...66R...1T} {66, R1}

\bibitem[\protect\citeauthoryear{{Taruya}, {Takada}, {Hamana}, {Kayo}  \&
  {Futamase}}{{Taruya} et~al.}{2002}]{Taruya02}
{Taruya} A.,  {Takada} M.,  {Hamana} T.,  {Kayo} I.,   {Futamase} T.,  2002,
  \mn@doi [\apj] {10.1086/340048}, \href
  {http://adsabs.harvard.edu/abs/2002ApJ...571..638T} {571, 638}

\bibitem[\protect\citeauthoryear{{Uhlemann}, {Codis}, {Pichon}, {Bernardeau}
  \& {Reimberg}}{{Uhlemann} et~al.}{2016}]{Uhlemann16log}
{Uhlemann} C.,  {Codis} S.,  {Pichon} C.,  {Bernardeau} F.,   {Reimberg} P.,
  2016, \mn@doi [\mnras] {10.1093/mnras/stw1074}, \href
  {http://adsabs.harvard.edu/abs/2016MNRAS.460.1529U} {460, 1529}

\bibitem[\protect\citeauthoryear{{Uhlemann} et~al.,}{{Uhlemann}
  et~al.}{2017a}]{Uhlemann17Bias}
{Uhlemann} C.,  et~al., 2017a, preprint, \href
  {http://adsabs.harvard.edu/abs/2017arXiv170508901U} {} (\mn@eprint {arXiv}
  {1705.08901})

\bibitem[\protect\citeauthoryear{{Uhlemann}, {Pajer}, {Pichon}, {Nishimichi},
  {Codis}  \& {Bernardeau}}{{Uhlemann} et~al.}{2017b}]{Uhlemann17nonG}
{Uhlemann} C.,  {Pajer} E.,  {Pichon} C.,  {Nishimichi} T.,  {Codis} S.,
  {Bernardeau} F.,  2017b, preprint, \href
  {http://adsabs.harvard.edu/abs/2017arXiv170802206U} {} (\mn@eprint {arXiv}
  {1708.02206})

\bibitem[\protect\citeauthoryear{{Uhlemann}, {Codis}, {Kim}, {Pichon},
  {Bernardeau}, {Pogosyan}, {Park}  \& {L'Huillier}}{{Uhlemann}
  et~al.}{2017c}]{Uhlemann17Kaiser}
{Uhlemann} C.,  {Codis} S.,  {Kim} J.,  {Pichon} C.,  {Bernardeau} F.,
  {Pogosyan} D.,  {Park} C.,   {L'Huillier} B.,  2017c, \mn@doi [\mnras]
  {10.1093/mnras/stw3221}, \href
  {http://adsabs.harvard.edu/abs/2017MNRAS.466.2067U} {466, 2067}

\bibitem[\protect\citeauthoryear{{Valageas}}{{Valageas}}{2002}]{2002A&A...382..412V}
{Valageas} P.,  2002, \mn@doi [\aap] {10.1051/0004-6361:20011663}, \href
  {http://adsabs.harvard.edu/cgi-bin/nph-bib_query?bibcode=2002A%26A...382..412V&db_key=AST}
  {382, 412}

\bibitem[\protect\citeauthoryear{{Vale} \& {Ostriker}}{{Vale} \&
  {Ostriker}}{2004}]{ValeOstriker04}
{Vale} A.,  {Ostriker} J.~P.,  2004, \mn@doi [\mnras]
  {10.1111/j.1365-2966.2004.08059.x}, \href
  {http://adsabs.harvard.edu/abs/2004MNRAS.353..189V} {353, 189}

\bibitem[\protect\citeauthoryear{{White}}{{White}}{2016}]{White16}
{White} M.,  2016, preprint, \href
  {http://adsabs.harvard.edu/abs/2016arXiv160908632W} {} (\mn@eprint {arXiv}
  {1609.08632})

\bibitem[\protect\citeauthoryear{{White} \& {Padmanabhan}}{{White} \&
  {Padmanabhan}}{2009}]{WhitePadmanabhan09}
{White} M.,  {Padmanabhan} N.,  2009, \mn@doi [\mnras]
  {10.1111/j.1365-2966.2009.14732.x}, \href
  {http://adsabs.harvard.edu/abs/2009MNRAS.395.2381W} {395, 2381}

\bibitem[\protect\citeauthoryear{{Wolk}, {McCracken}, {Colombi}, {Fry},
  {Kilbinger}, {Hudelot}, {Mellier}  \& {Ilbert}}{{Wolk}
  et~al.}{2013}]{2013MNRAS.435}
{Wolk} M.,  {McCracken} H.~J.,  {Colombi} S.,  {Fry} J.~N.,  {Kilbinger} M.,
  {Hudelot} P.,  {Mellier} Y.,   {Ilbert} O.,  2013, \mn@doi [\mnras]
  {10.1093/mnras/stt1111}, \href
  {http://cdsads.u-strasbg.fr/abs/2013MNRAS.435....2W} {435, 2}

\bibitem[\protect\citeauthoryear{{Xavier}, {Abdalla}  \& {Joachimi}}{{Xavier}
  et~al.}{2016}]{Xavier16}
{Xavier} H.~S.,  {Abdalla} F.~B.,   {Joachimi} B.,  2016, \mn@doi [\mnras]
  {10.1093/mnras/stw874}, \href
  {http://adsabs.harvard.edu/abs/2016MNRAS.459.3693X} {459, 3693}

\bibitem[\protect\citeauthoryear{{Yang} \& {Saslaw}}{{Yang} \&
  {Saslaw}}{2011}]{YangSaslaw11SDSS}
{Yang} A.,  {Saslaw} W.~C.,  2011, \mn@doi [\apj]
  {10.1088/0004-637X/729/2/123}, \href
  {http://adsabs.harvard.edu/abs/2011ApJ...729..123Y} {729, 123}

\bibitem[\protect\citeauthoryear{{Yoshisato}, {Matsubara}  \&
  {Morikawa}}{{Yoshisato} et~al.}{1998}]{Yoshisato98}
{Yoshisato} A.,  {Matsubara} T.,   {Morikawa} M.,  1998, \mn@doi [\apj]
  {10.1086/305534}, \href {http://adsabs.harvard.edu/abs/1998ApJ...498...48Y}
  {498, 48}

\bibitem[\protect\citeauthoryear{{Yoshisato}, {Morikawa}, {Gouda}  \&
  {Mouri}}{{Yoshisato} et~al.}{2006}]{Yoshisato06}
{Yoshisato} A.,  {Morikawa} M.,  {Gouda} N.,   {Mouri} H.,  2006, \mn@doi
  [\apj] {10.1086/498496}, \href
  {http://adsabs.harvard.edu/abs/2006ApJ...637..555Y} {637, 555}

\bibitem[\protect\citeauthoryear{{de Jong} et~al.,}{{de Jong}
  et~al.}{2013}]{KIDS}
{de Jong} J.~T.~A.,  et~al., 2013, The Messenger, \href
  {http://adsabs.harvard.edu/abs/2013Msngr.154...44J} {154, 44}

\makeatother
\end{thebibliography}
\appendix

\section{Depth, collapse and cumulants} 
\label{sec:details}
\subsection{Leading order cumulants depending on shape}
\label{app:skewness}
While a closed-form spherical collapse solution is known for densities in spheres and densities in infinitely long cylinders (that correspond to two-dimensional disks on the sky), no solution for finite cylinders is known yet. To get an intuition about the finite-size effects related to cylinders, let us perturbatively compute the cumulants at leading order. Once the second order kernel from standard (Eulerian) perturbation theory  is written as \citep[see e.g.][]{Bernardeau02LSS}
\begin{align}
F_2(\vk_1,\vk_2) = \frac{5}{7} + \frac{1}{2} \frac{(\vk_1\cdot\vk_2)(k_1^2+k_2^2)}{k_1^2k_2^2} + \frac{2}{7} \frac{(\vk_1\cdot\vk_2)^2}{k_1^2k_2^2}\,,
\end{align}
the leading-order skewness for a smoothing kernel of arbitrary shape can be computed as
\begin{align}
\label{eq:skewness}
\kappa_3(\mR)&=3\, \langle (\delta^{(1)})^2 \delta^{(2)}\rangle \\
&= 6 \int \frac{\dd^3k_1}{(2\pi)^3}\! \int \frac{\dd^3k_2}{(2\pi)^3} F_2(\vk_1,\vk_2)P_L(k_1)P_L(k_2) \\
\notag &\qquad \qquad\qquad W(\vk_1,\mR)W(\vk_2,\mR)W(-(\vk_1+\vk_2),\mR)\,.
\end{align}
The reduced skewness $S_3={\kappa_3}/{\sigma^4}$ is obtained as the ratio of the skewness and the square of the variance which can be evaluated in linear theory according to equation~\eqref{eq:covmatrix}. The result of a numerical integration for the tree-order skewness with a finite cylindrical filter is shown in Fig.~\ref{fig:S3Horizon}.
 \begin{figure}
\includegraphics[width=0.9\columnwidth]{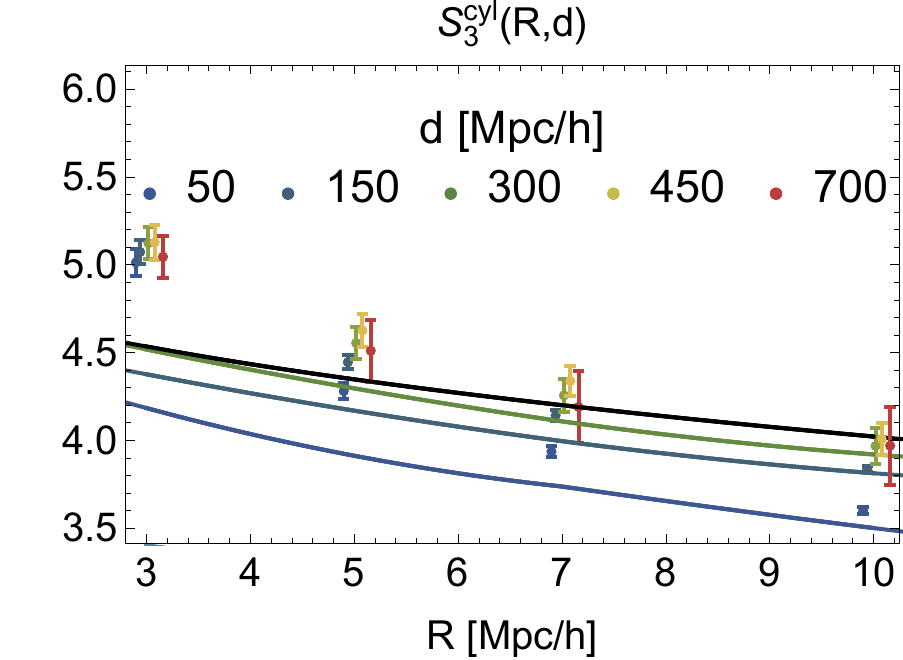}
\caption{Skewness of the density in cylindrical cells with length $d=50,150,300,450,700$ Mpc/$h$ (lower blue to higher red) and radii $R=3,5,7,10$ Mpc/$h$ at redshift $z=0.7$ from the HR4 measurements (data points with error bars) and the tree-order skewness computed from equation~\eqref{eq:skewness} for $d=50,150,300$ as well as the infinitely long cylinder limit according to equation~\eqref{eq:smoothingeffects2D} (black line). }
   \label{fig:S3Horizon}
\end{figure}
{\it Analytical computation of cumulants from symmetry.} When computing the cumulants for unsmoothed fields (hence $W=1$), one  has to compute an angular average of the $F_2$ kernel which gives for spherical symmetry and in an EdS universe the well-known result $S_3^{\rm 3D}=34/7$. Using the approximate power law for spherical collapse from equation~\eqref{eq:spherical-collapse} can be matched to $S_3=3(1+1/\nu)$ giving $\nu_{\rm 3D}=21/13$.
In analogy, one obtains $S_3^{\rm 2D}=36/7$ for the disk symmetry which holds for an infinite cylinder where on can switch to cylindrical coordinates where $\vk=(\vk_\perp,k_\parallel)$ and $0\simeq k_\parallel \ll k_\perp$ and hence $\nu_{\rm 2D}=7/5=1.4$. 
When including the top-hat filtering in real space, the effects of smoothing for a power law initial spectrum $P(k)\propto k^n$ are given in \cite{Bernardeau94smooth,Bernardeau1995} and read
\begin{align}
\label{eq:smoothingeffects}
S_3^{\rm 3D} &=\frac{34}{7}+ \gamma_1 \simeq \frac{34}{7} -(n_{\rm 3D}+3) \,, \quad \gamma_1=\frac{\dd\log\sigma^2_{\text{sph}}}{\dd\log R}\,,\\
\label{eq:smoothingeffects2D}
S_3^{\rm 2D} &\simeq \frac{36}{7} -\frac{3}{2}(n_{\rm 2D}+2)
\end{align}
There is a slight $\Omega_m$ dependence in the constant term which is bracketed by $21/4> S_3^{\rm 2D} >36/7$ for $\Omega_m\in [0,1]$.

For finite-size cylinders, one needs to take the component of the wave-vector parallel to the line of sight into account such that one cannot obtain a direct analytical expression but has to numerically evaluate equation~\eqref{eq:skewness}.
 Fig.~\ref{fig:S3Horizon} shows the reduced skewness measured from the HR4 simulation. As expected by cylindrical collapse, the reduced skewness becomes independent of depth if the axis ratio is large enough (typically for $2R/d\gtrsim 1/10$). Note that the nonlinear variance $\sigma^2(R,d)\propto 1/d$, see equation~(31) in \cite{BV2000}, such that the product $\sigma^2(R,d)\times d$ remains constant to a very good accuracy.
 
 \subsection{Large deviation statistics with varying cylinder depth}
\label{app:LDS}
 Fig.~\ref{fig:saddlePDFvsHorizonvaryD}  compares the theoretical prediction for infinitely long cylinders, hence effectively 2D-spherical densities when fixing the radius $R$ and varying the depth $d$. The limit of essentially infinitely long cylinders is achieved at 10\% accuracy in the PDF already for a depth to radius ratio of about $(2R)/d\simeq 1/5$ and quickly improves with increasing depth. This effect can be also seen in Fig.~\ref{fig:S3Horizon}  which compares the measured skewness to the perturbation theory result for the corresponding hierarchical clustering ratio.
\begin{figure}
\includegraphics[width=\columnwidth]{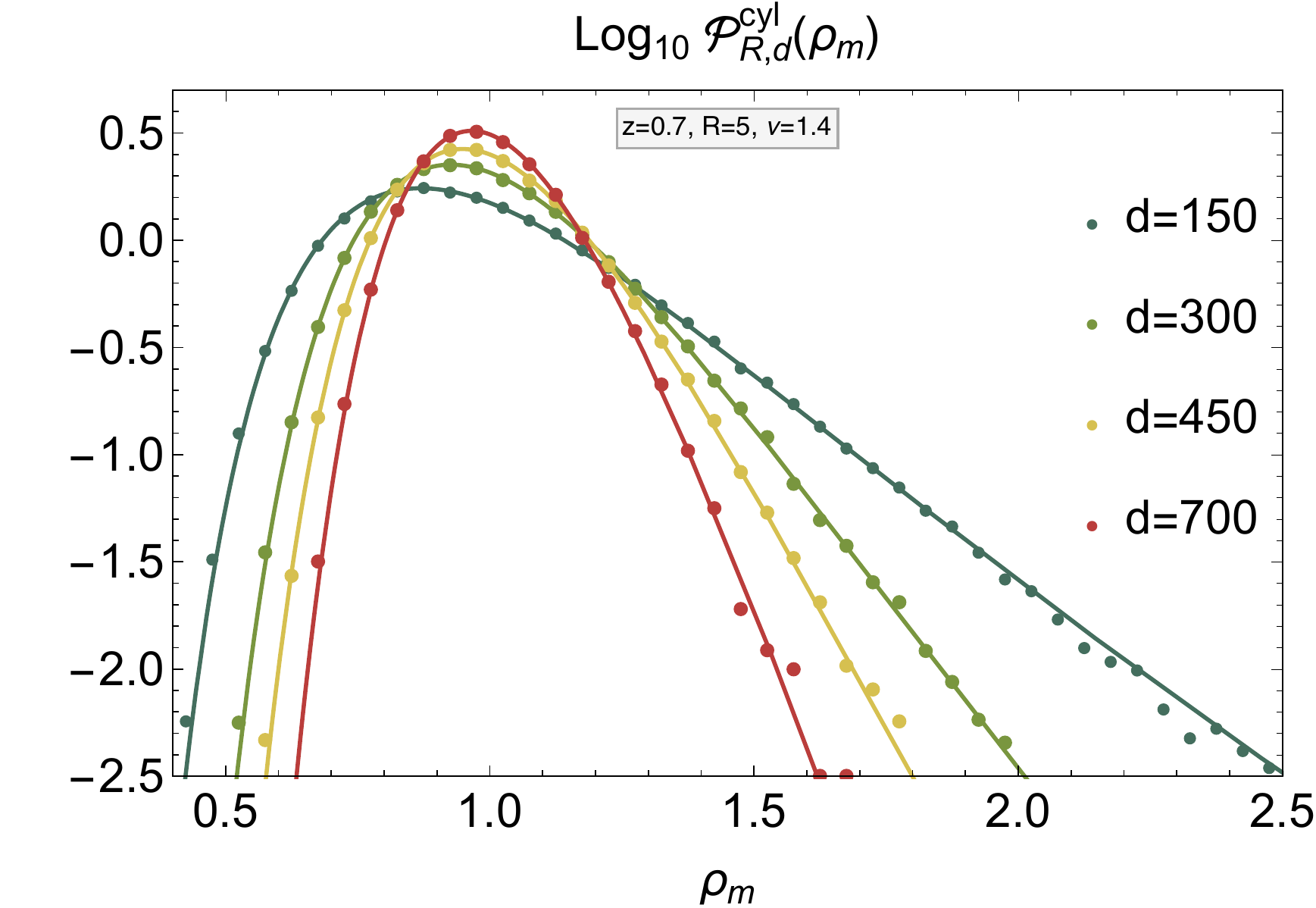}\\
\includegraphics[width=\columnwidth]{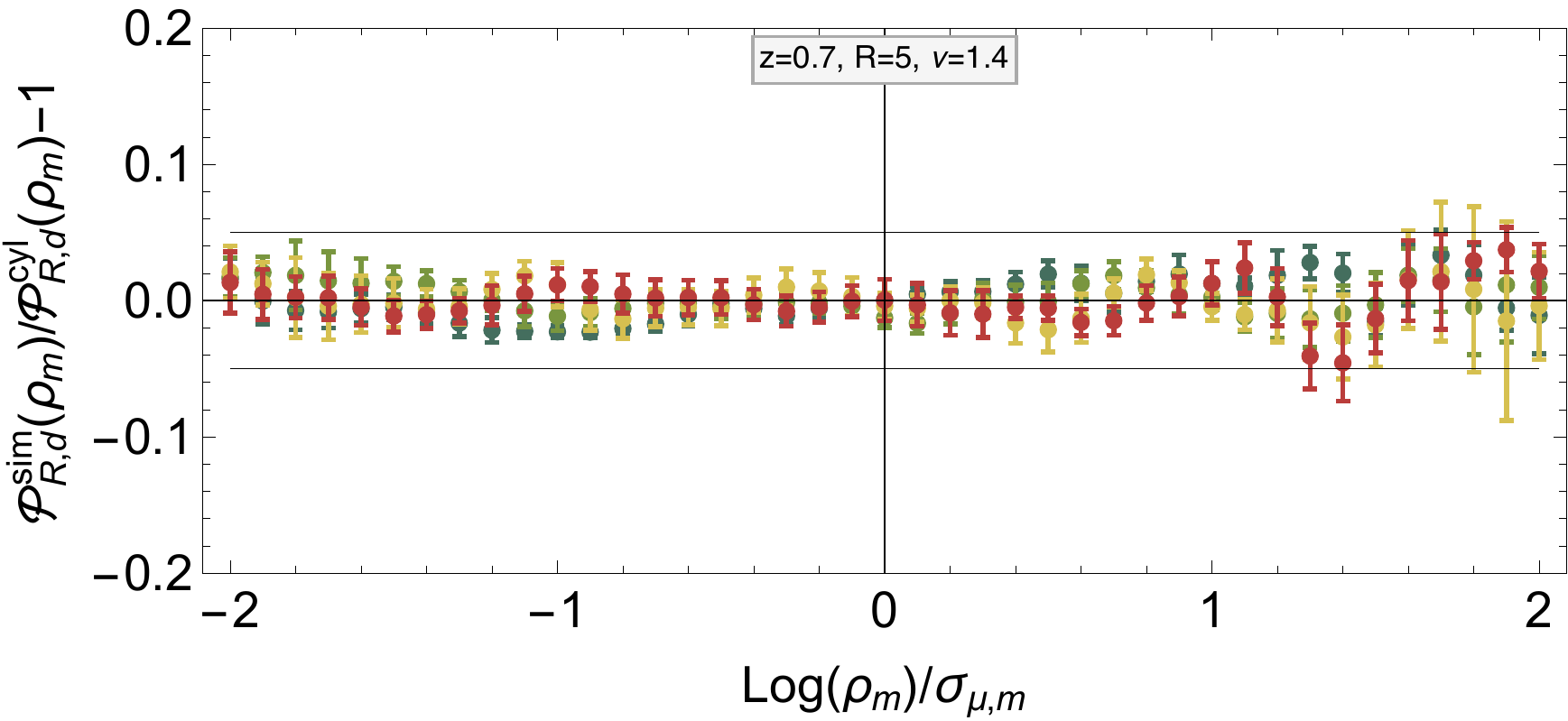}
\caption{{\it (Upper panel)} PDF of the projected density in cylindrical cells of length $d=150,300,450,700$ Mpc/$h$ and radius $R=5$ Mpc/$h$ at redshift $z=0.7$. Shown is a comparison of HR4 measurements (data points) and the prediction computed with the   cylindrical filter~\eqref{eq:filtercyl} with mass-splitting for long cylinders \eqref{eq:mapdisk} together with the spherical collapse parametrized by $\nu=1.4$ (solid lines).
{\it (Lower panel)} Residuals of the theoretical prediction against the HR4 measurements with error bars.}
   \label{fig:saddlePDFvsHorizonvaryD}
\end{figure}
\subsection{Scale-dependence of the linear covariance}
\label{app:PSparam}
The scale-dependence of the linear variance from equation~\eqref{eq:covmatrix} for varying radius and depth of the cylinder is shown in Fig.~\ref{fig:scaledepvariance} and compared to the variance computed with a spherical filter of equal volume. 
\begin{figure}
\includegraphics[width=0.85\columnwidth]{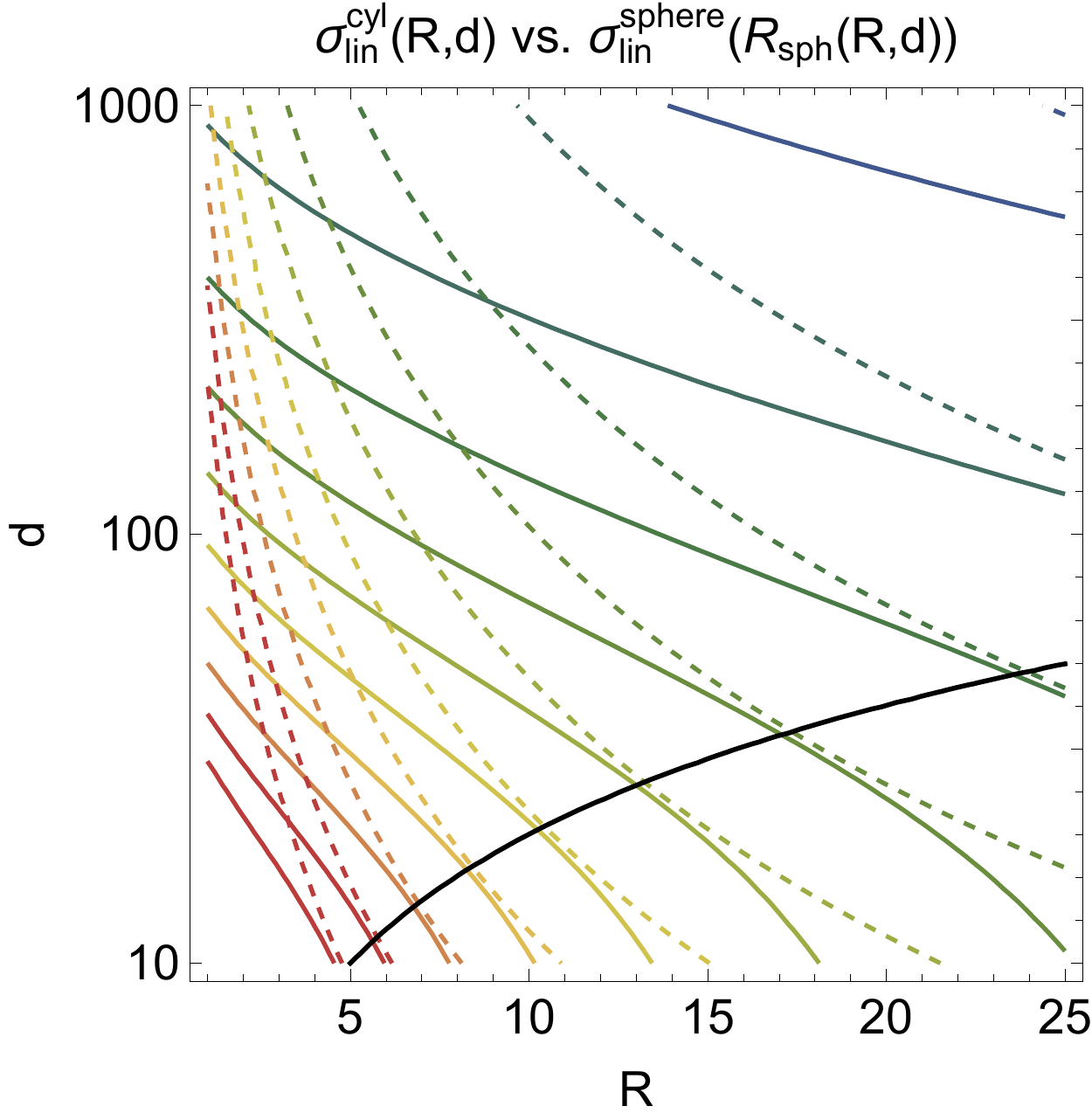}
 \caption{Scale dependence of the linear variance $\sigma$ induced by a cylindrical filter with radius $R$ and depth $d$, both in Mpc$/h$. The contours indicate variances from 0.1 (blue) to 1.0 (red) in steps of 0.1. The cylindrical filter (solid lines) is compared to a volume-equivalent spherical filter with $R_{\rm sph}=(3/4 R^2 d)^{1/3}$ (dashed lines). As expected, they are  approximately equal for an axis ratio $2R/d=1$ (black line).}
  \label{fig:scaledepvariance}
\end{figure}
For concentric cylinders of different radii $R_i$ but identical depth $d$, for the sake of simplicity, the covariance matrix from equation~\eqref{eq:covmatrix} is  parametrized in analogy to a power-law initial spectrum with spectral index $n_{\rm 2D}(R_p) \simeq n_{\text{cyl}}(R_p,d)$ by 
\begin{subequations}
\label{eq:sigijparam}
\begin{align}
\frac{\sigma^{2}(R_{i},R_{i},d)}{\sigma^2(R_{p},d)}&=\left(\frac{R_{i}}{R_{p}}\right)^{-n_{\text{cyl}}(R_{p},d)-2}\,,\\
\frac{\sigma^{2}(R_{i},R_{j> i},d)}{\sigma^2(R_{p},d)}
&=\,{\cal G}\left(\frac{R_i}{R_p},\frac{R_j}{R_p},n_{\text{cyl}}(R_{p},d)\right)\,,
\end{align}
\end{subequations}
where the 2D-spectral index is obtained as
 \begin{align} 
\label{eq:ncyl}
n_{\text{cyl}}(R,d)=-2-\frac{d\log\sigma^2_{\text{cyl}}(R,d)}{d\log R}\,.
\end{align}
and
\begin{align}
{\cal G}(x,y,n)&=\frac{\displaystyle \int{\dd^{3}k\,}k^{n}W_{\rm 2D}(k x)W_{\rm 2D}(k y)}{\displaystyle\int{\dd^{3}k\,}k^{n}W_{\rm 2D}(k R_{p})W_{\rm 2D}(k R_{p})}\,,\nonumber
\\
&=
  \! \frac{ (x\!+\!y)^{\alpha} \!\! \left(\!x^2\!+\!y^2\!-\!\alpha x y\right)\!-\!(y\!-\!x)^{\alpha} 
 \!  \!\left(\!x^2\!+\!y^2\!+\!\alpha x y\right)}
   {2^{\alpha}(n+1) x^3 y^3  },\nonumber
\end{align}
with $\alpha=1-n$. The key parameter in the prediction of the PDF is the value of the variance at the pivot radius $R_p$ and depth $d$ which is measured in the simulation and use as an input to our theoretical model.

\subsection{Joint PDF of dark matter around halo density}
Figs.~\ref{fig:saddlePDF2cellhmvsmeasurements}~and~\ref{fig:saddlePDFrho12hmvsmeasurements} show the joint PDF of matter density in rings around halo density in cylinders and the conditional PDF of matter density in rings given tracer density. The curves of the theoretical prediction for the conditional look almost identical to Fig.~\ref{fig:saddlePDFrho12vsmeasurements}, which is due to the fact that the best-fit bias model leads to a similar split $\rho_m(\rho_h=1)\simeq 1.02$ between over- and underdensity for both matter and haloes. A possible cause of the residuals seen in the PDFs conditional-on-halo-density is the scatter around the mean bias relation which should increase the probability of unlikely events such as over-/underdense matter rings around under-/overdense halo cores that are not captured in a one-to-one bias model.
\begin{figure}
\includegraphics[width=\columnwidth]{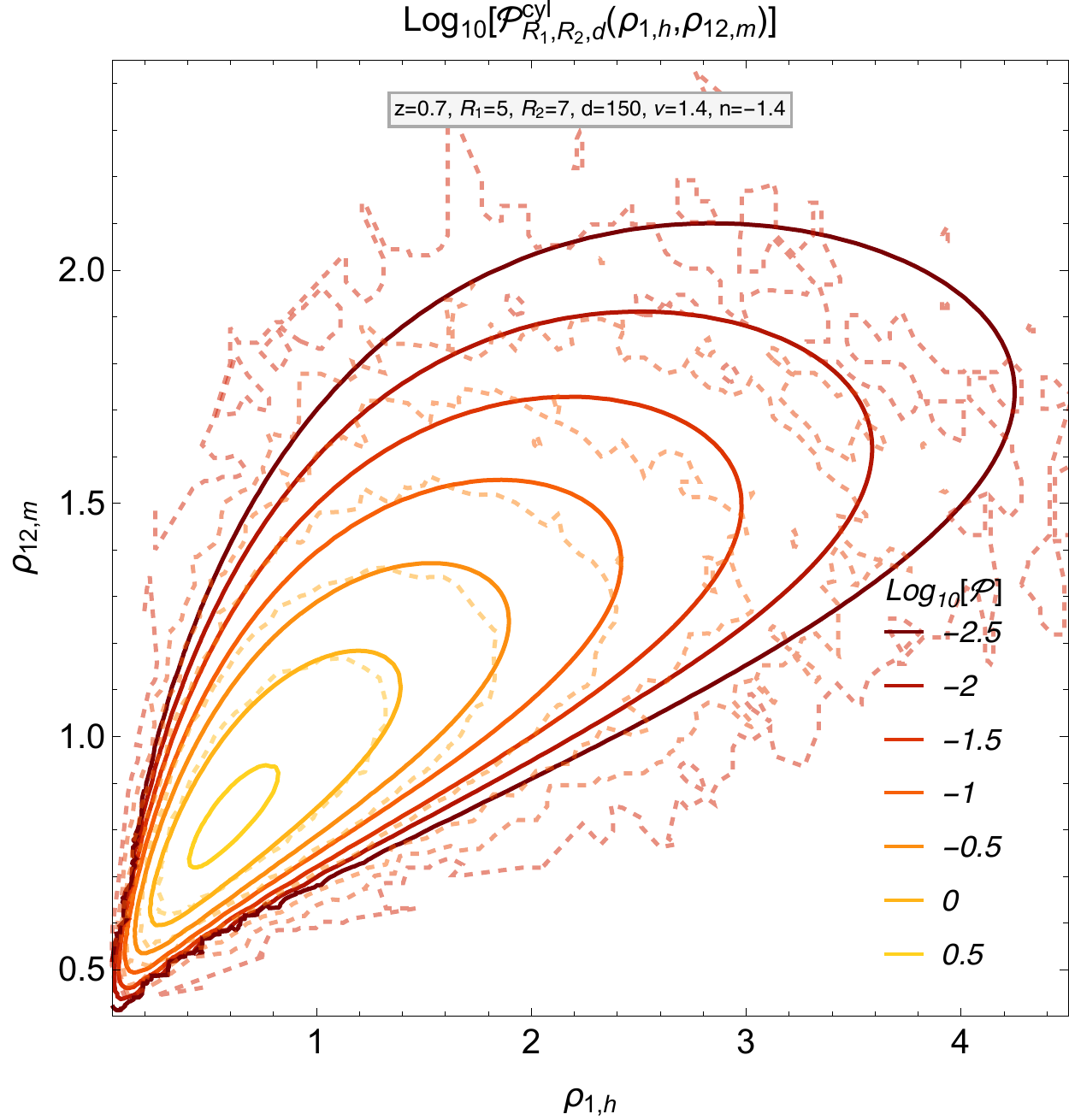}
\caption{Joint PDF of the projected halo densities in a cylinder of depth $d=150$ Mpc/$h$ and radius $R_1=5$ Mpc/$h$ and dark matter density in a surrounding cylindrical ring with $R_2=7$ Mpc/$h$ at redshift $z=0.7$ from the saddle point approximation with $\nu=1.4$ and the disk filter for a power-law spectrum with $n=-1.4$ and the best fit bias parameters from Table~\ref{tab:biasfit} (solid lines) compared to the HR4 measurements (dashed lines).
}
\label{fig:saddlePDF2cellhmvsmeasurements}
\end{figure}
\begin{figure}
\includegraphics[width=\columnwidth]{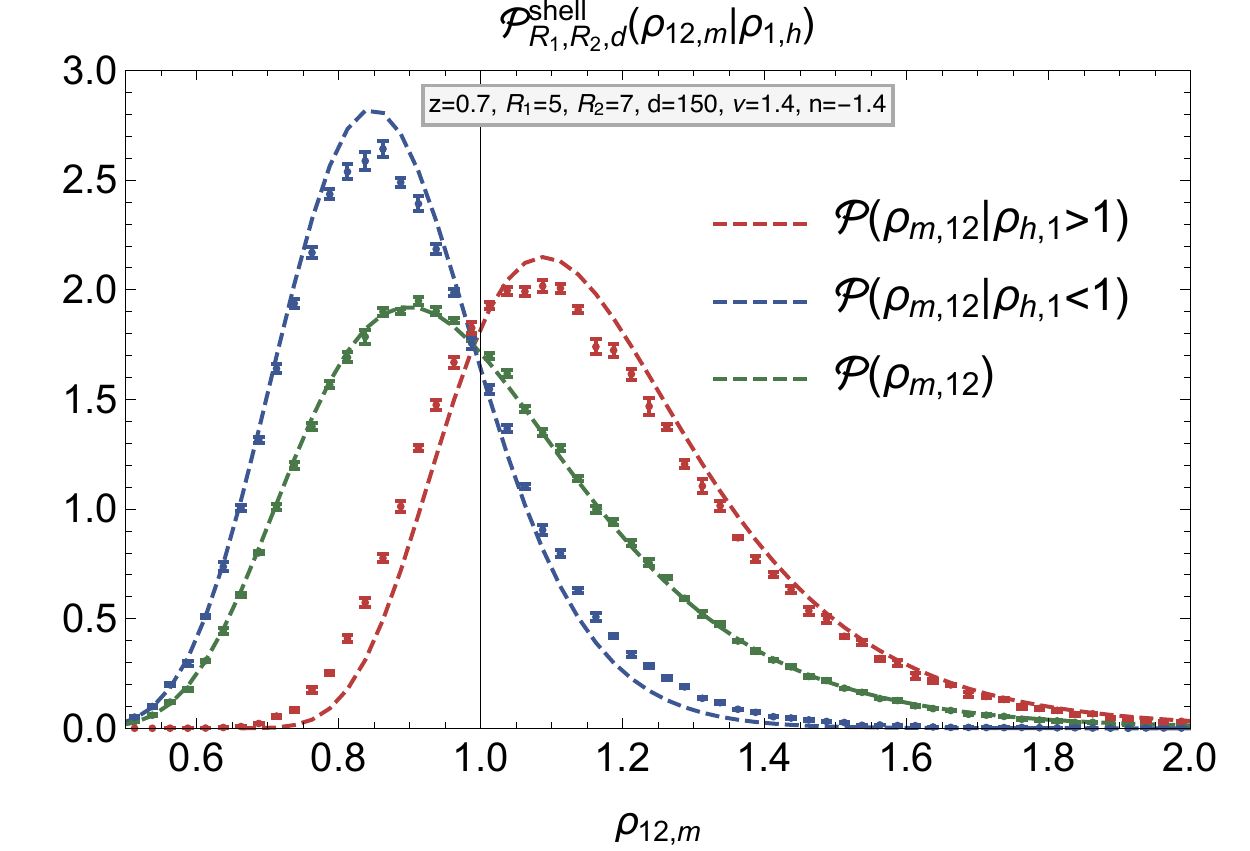}
\caption{Marginalised PDF of the matter densities in cylindrical rings of depth $d=150$ Mpc/$h$ and radii $R_1=5$ Mpc/$h$ and $R_2=7$ Mpc/$h$ at redshift $z=0.7$ from the saddle point approximation with $\nu=1.4$ and the disk filter for a power-law spectrum with $n=-1.4$ together with the best fit bias model from Table~\ref{tab:biasfit} for arbitrary inner tracer densities (dashed, green line), overdensities in tracers (dashed red line) and underdensities in tracers (dashed blue line) compared to the HR4 measurements (data points). Comparing the theoretical prediction to the PDF given over- or underdensity in dark matter in Fig.~\ref{fig:saddlePDFrho12vsmeasurements}, they look virtually identical.}
   \label{fig:saddlePDFrho12hmvsmeasurements}
\end{figure}
\section{Lognormal reconstruction}
\label{app:lognormal}
One widespread phenomenological ansatz for the one-point PDF of either dark matter or its tracers is the lognormal distribution \citep{ColesJones91}. Assuming a lognormal density field, the one-point PDF is fully determined by the variance of the log-density and hence one can attempt a reconstruction of the density PDF using the functional form
\begin{align}
\label{eq:lognormal}
\mP_{\rm LN}(\rho\,|\,\sigma_\mu)&=\frac{1}{\sqrt{2\pi\sigma_\mu}} \frac{1}{\rho} \exp\left[-\frac{(\log\rho +  \sigma_\mu^2/2)^2}{2 \sigma_\mu^2}\right]\,.
\end{align} 
Fig.~\ref{fig:lognormalPDFvsHorizonfixd}  shows a comparison of the lognormal model for the PDF of dark matter in cylinders with fixed depth and varying radius. 
When plotted as ratio between the measured PDF and the lognormal reconstruction, a clear residual skewness is visible which leads to deviations at the 5\%-20\% level for all densities of interest. Overall, the accuracy improves with increasing cylinder radius and cylinder depth.
\begin{figure}
\includegraphics[width=\columnwidth]{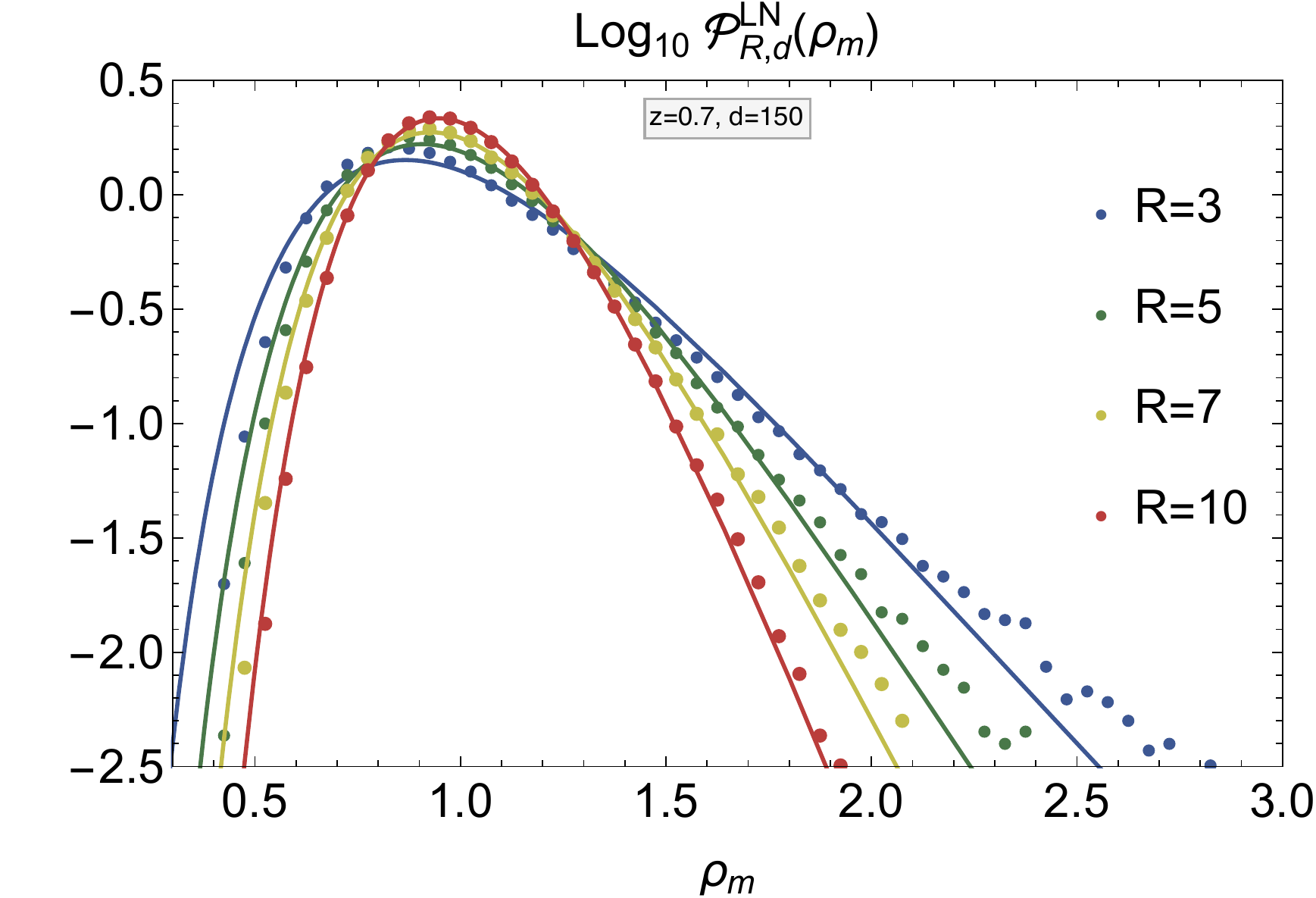}
\includegraphics[width=\columnwidth]{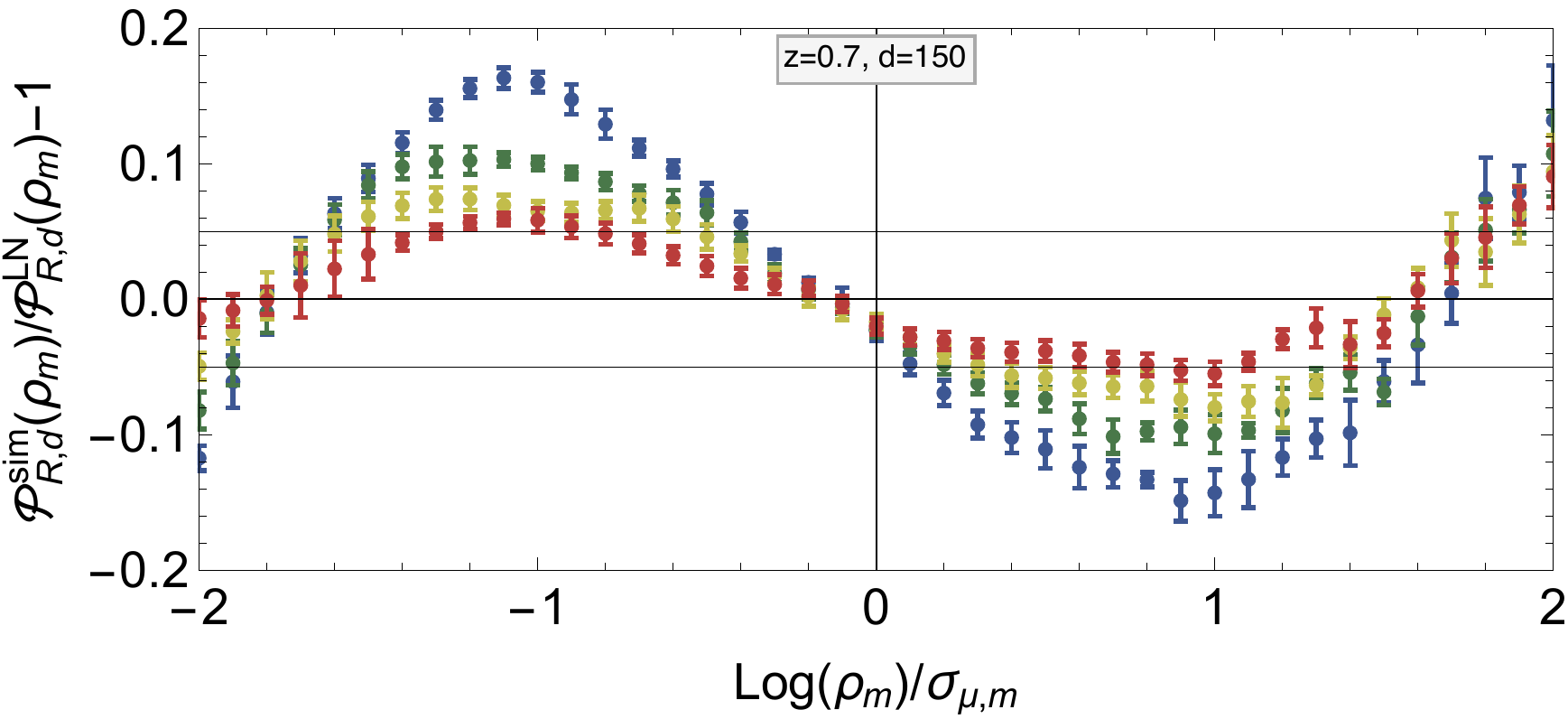}
\caption{Comparison of lognormal PDF model from equation~\eqref{eq:lognormal} with measured log-variance as given in Table~\ref{tab:variance} for fixed depth $d=150$ Mpc$/h$ and varying radius against the measured PDF for dark matter. This plot is the analogue of Fig.~\ref{fig:saddlePDFvsHorizoncyl} for the lognormal reconstruction.}
   \label{fig:lognormalPDFvsHorizonfixd}
\end{figure}

\end{document}